\documentclass[aps,pra,reprint,amsmath,amssymb,
superscriptaddress,showpacs,showkeys,nofootinbib]{revtex4-2}
\usepackage{amsmath}    
\usepackage{graphicx}   
\usepackage{verbatim}   
\usepackage{color}      
\usepackage{subfigure}  
\usepackage{hyperref}   
\usepackage[warn]{mathtext}    
\usepackage{dcolumn}    
\usepackage{bm}         
\usepackage{soul} 
\definecolor{vs}{rgb}{0.1,0.4,0.1}              

\usepackage{hyperref}
\usepackage{verbatim}
\newcommand\wordcount{\verbatiminput{\jobname.sum}}

\begin{document}

\begin{minipage}{\textwidth}
{\small ISSN 0021-3640, JETP Letters, 2023, {\bf 118}, No. 6, pp. 414–425 (2023).\\
Russian Text published in Pis’ma v Zhurnal Eksperimental’noi i Teoreticheskoi Fiziki, {\bf 118}, No. 6, pp. 417–429 (2023).\\
\mbox{}}
\end{minipage}
\title{Effect of Energy Conservation Law, Space Dimension, and Problem Symmetry\\ on the Poynting Vector Field Singularities}


\author{M.I. Tribelsky  (ORCID: 0000-0002-4169-6740) }
\email[]{\mbox{E-mail: mitribel\_at\_gmail.com \@(replace ``\_at\_" by @)}
}
\homepage[]{\\https://polly.phys.msu.ru/en/labs/Tribelsky/}
\affiliation{
M. V. Lomonosov Moscow State University, Moscow, 119991, Russia}
\affiliation{Center for Photonics and 2D Materials, Moscow Institute of Physics and Technology, Dolgoprudny 141700, Russia}




\begin{abstract}
A brief review is given of the author’s recent achievements in classifying singular points of the Poynting vector patterns in electromagnetic fields of complex configuration. The deep connection between the topological structure of the force lines pattern and the law of energy conservation, the symmetry of the problem, and the dimension of the space has been unveiled.\\
\mbox{}\\
{\bf DOI:} 10.1134/S0021364023601859
\end{abstract}


\maketitle
\section{Introduction\label{sec:introduc}}
Singular points of electromagnetic field significantly impact its topological structure. The type and location of these points are crucial to the overall structure of the entire field pattern. For this reason, many researchers are interested in studying this topic. Recent advances in experimental setups for generating and analyzing complex light beam configurations have increased attention to the study of this issue; see, e.g.,~\cite{Mokhun2007, Novitsky2009,Dennis2009,Gao2014,Yue2019,Angelsky2021} and references therein. 

The above is fully applied to the Poynting vector field  $\mathbf{S}(\mathbf{r})$. In particular, the singularities of this field are essential in resonance scattering of light by sub-wavelength objects. For example, in light scattering by nanoparticles, the topological structure of the Poynting vector field determines the complex energy circulation in the vicinity of and within the light-scattering particle, while the divergence of this field determines the dissipation of electromagnetic energy.

In addition to the purely academic interest, these issues are of great practical importance for various nanotechnologies, as nanoscale controlled energy release is a unique way to achieve a localized effect on various materials. The Poynting vector field also contributes to the ponderomotive forces~\cite{Gao2014}, which is essential for manipulating nanoobjects using electromagnetic field; see, e.g., review~~\cite{Gao2017} for details.

At the same time, experimental measurement of the Poynting vector field is a very difficult task~\cite{Mokhun2012}. Moreover, the measurements are hardly possible in the practically important case of the field inside the solid-state scattering object. In such a situation, the theoretical description of this field is the only tool for its study.

Since the pioneering work of Bohren~\cite{Bohren1983} many authors have carried out such study~\cite{Wang2004,Bashevoy2005,tribelsky2006anomalousPRL,luk2007peculiaritiesCOLA,luk2007peculiaritiesJOA,CanosValero2021,Tribel:2022_UFN,Yue2022,Geints2022,Geints2022a}. In particular, work~\cite{Novitsky2009} presents a detailed classification of the Poynting vector field singularities.

However, most studies relate to examining features of complex configuration fields. Many interesting, important, and often unexpected results have been obtained in these studies. They include the singularities of nonparaxial Bessel beams~\cite{Gao2014}; singularities in superoscillating fields~ \cite{Berry2019}; singularities due to interference with toroidal modes~\cite{ospanova2021generalized}; the appearance of a reverse (relative to the incident wave) energy flow near the axis of the sharply focused vortex beam~\cite{kotliar2018obratnyi, Kotlyar2019}, etc. As a rule, the theoretical description of these features implies cumbersome computations involving such concepts as topological charges~\cite{Berry2004_TC,kotlyiarBookTC}, geometric Pancharatnam–Berry phase~\cite{berry1984quantal,Cohen2019}, etc.

Meanwhile, the most straightforward description in terms of the traditional Poincar\'{e} classification (saddle, node, etc.) of the spontaneously emerging singularities at the scattering of a plane, linearly polarized electromagnetic wave does not require cumbersome calculations and the introduction of new concepts. Nevertheless, it still needs to be completed. In particular, the fact that the field of the Poynting vector satisfies the energy conservation law, which, together with the dimension of space and the symmetry of the problem, imposes significant constraints on both the type of the singular points and the bifurcation scenario of their formation (annihilation), has not yet received sufficient attention. The author’s recent results~\cite{Tribelsky2022,Tribelsky2022_Nanomat_diss,Tribel:2023_arXiv}, briefly discussed in the present paper, have partly filled up this gap. The mini-review format does not allow other aspects of the problem to be discussed in detail here. The reader interested in them is addressed to other works cited in this paper.

Note also that in terms of the theoretical description of the problem, the solution of Maxwell’s equations defines the fields $\mathbf{E}$ and $\mathbf{H}$. As for the field $\mathbf{S}$, it is {\it calculated\/} based on the obtained fields $\mathbf{E}$ and $\mathbf{H}$, and, in this sense, it is secondary. As we will see later, this fact plays an essential role in understanding the physical meaning of the singularities of such a field.

The paper has the following structure: Sec.~\ref{sec:formulation} gives the problem formulation; Sec.~\ref{sec:nodiss} deals with singularities in a non-dissipative environment based on the exact solutions for a sphere (\ref{sec:Sphere}) and a cylinder (\ref{sec:cyl}); Sec.~\ref{sec:diss} discusses the dissipative effects; Sec.~\ref{sec:oblique} focuses on the effects of controlled symmetry breaking; Sec.~\ref{sec:concl} formulates some unsolved problems, contains conclusions and acknowledgments. 

\section{Problem formulation\label{sec:formulation}}

It is well known that the local topological properties of a given singularity determine the field's behavior in its neighborhood. The global behavior of this field at a distance only weakly affects it. For this reason, the phenomenological theory discussed below is universal. It is insensitive to changes in the scattered beam's geometry, its shape, and the optical properties of the scattering object. However, concrete examples illustrating this theory used below are based on calculations with the help of the exact solutions of the corresponding problems of scattering of a plane linearly polarized electromagnetic wave by a sphere of radius $R$ or an infinite right circular cylinder of cross-sectional radius $R$. These solutions are well known; see, e.g., ~\cite{Bohren1998}. Such an approach made it relatively easy to find the singular points themselves and to control the calculations' accuracy to be sufficient for a safe resolution of all the necessary details of the phenomena under discussion.
   
The magnetic permeability of the scattering body $\mu$ is assumed to be equal to unity (which corresponds to optical frequencies); and its permittivity $\varepsilon = \varepsilon'+i\varepsilon'' = const$, where $\varepsilon''>0$. If necessary, the results obtained can be easily generalized to the case of $\mu$ other than unity and/or $\varepsilon''<0$ (active scattering body with population inversion).

In the theoretical description of scattering problems, the field outside the scattering object is usually presented as the sum of the incident wave field and the radiation field scattered by the object. To avoid misunderstandings, we emphasize that anywhere in the present paper, the fields $\mathbf{E}$ and $\mathbf{H}$ mean the full fields equal to the indicated sum.

The Poynting vector is assumed to be real. It is defined in the usual way~\cite{Landau_Electrodyn}:
\begin{equation}\label{eq:Poynting_EH}
  \mathbf{S}=\frac{c}{16\pi}(\mathbf{[E^*H]} +\mathbf{[EH^*]}),
\end{equation}
where $c$ is the speed of light in a vacuum, and the asterisk designates complex conjugation.
Note, however, that in some cases it is expedient to introduce the complex Poynting vector \mbox{$\hat{\mathbf{S}}=\frac{c}{8\pi}\mathbf{[E^*H]}$,} whose imaginary part has the meaning of an oscillating stored energy flow~\cite{Jackson1998}. This quantity plays a vital role in some problems of the interaction of light with matter~\cite{bliokh2014magnetoelectric,Bliokh2014,Bekshaev2015,Xu2019,Khonina2021,Tang2010,Lininger2022}. As a rule, the results discussed below, obtained for real $\mathbf{S}$, can easily be generalized to the case of complex $\hat{\mathbf{S}}$.

The pattern of the Poynting vector field consists of a set of force lines, which, by analogy with hydrodynamics, we will also call streamlines. The tangents to these lines at each point correspond to the direction of the $\mathbf{S}$ vector at this point. It is convenient to specify the streamline equation in the parametric form $\mathbf{r} = \mathbf{r}(t)$. Here, $t$ is a dimensionless parameter, generally speaking, in no way connected with the actual time. With this representation, the ``velocity'' $d\mathbf{r}/dt$ will be directed tangentially to the streamline, i.e., proportional to the vector $\mathbf{S}(\mathbf{r})$. By properly scaling $t$, we can always set the corresponding proportionality factor to unity. Then, the equation that determines the streamlines takes the form:
\begin{equation}\label{eq:stream_vector}
  \frac{d\mathbf{r}}{dt} = \mathbf{S}(\mathbf{r}).
\end{equation}

Next, we introduce dimensionless variables, normalizing the coordinate $\mathbf{r}$ to $R$, and the fields $\mathbf{E}, \mathbf{H}$, and $\mathbf{S}$ to the corresponding values in the incident wave. Since only dimensionless variables are in use below, we will keep the same notations for them since this cannot lead to misunderstandings.

The vector $\mathbf{S}(\mathbf{r})$ singularities, as well as those of any vector field, correspond to the intersection points of its lines of force. Since, as already noted, the tangents to the lines of force determine the direction of the vector $ \mathbf{S}(\mathbf{r})$, their intersection at some point $\mathbf{r}_s$ means that the vector $\mathbf{ S}(\mathbf{r}_s)$ is not uniquely defined. On the other hand, the fundamental physical principles require that the field $ \mathbf{S}(\mathbf{r})$ must be uniquely defined. The only way to reconcile these two facts is to suppose that $\mathbf{S}(\mathbf{r}_s) = 0$. 

Obviously, there are three possibilities for this condition to fulfill: \mbox{(a) $\mathbf{E}(\mathbf{r}_s)=0$,} (b) $\mathbf{H}(\mathbf{r}_s)=0$ and \mbox{(c) $\mathbf{S}(\mathbf{r}_s)=0$,} when neither (a) nor (b) hold (this case corresponds to the formation of a standing wave). In accordance with the terminology introduced in Ref.~\cite{Novitsky2009}, we will call such singular points {\it E-induced\/}, {\it H-induced\/} and {\it polarization-induced}, respectively. Calling the first two types both {\it field-induced} is also convenient. We will call this classification electromagnetic to distinguish it from the standard Poincar\'{e} classification (node, focus, etc.).

It is important to emphasize that if in field-induced singularities, the entire vector product $\mathbf{[E^*H]}$ equals zero, the polarization-induced singularities require only its real part to vanish. As a result, field-induced singularities for the real and imaginary parts of the complex Poynting vector $\hat{\mathbf{S}}$ coincide, but generally speaking, polarization-type singularities do not.

\section{Non-dissipative limit\label{sec:nodiss}} 
\vspace*{10pt}
Note that the standard problem formulation for light scattering by a material object, in which the incident wave comes from infinity and the scattered one goes to infinity~\cite{Bohren1998}, is physically adequate only if this object is embedded in a non-dissipative medium, a particular case of which is a vacuum. Otherwise, the incident wave attenuates before reaching the scattering object. 

For this reason, all singularities outside the scattering body must be considered in a non-dissipative medium. In other words, the absence of dissipation in the vicinity of the singularity is essential for describing scattering by any, even strongly absorbing, particle. Since the scattering problem for a particle in a medium with an arbitrary value of $\varepsilon >0$ is reduced to the case of scattering in a vacuum by a trivial scale transformation~\cite{Bohren1998}, without losing the generality of the consideration, we can assume that the permittivity of the environment is equal to unity, which will be supposed below.

\subsection{Sphere\label{sec:Sphere}}

We begin our analysis of singularities by discussing the case of light scattering by a finite object with a plane of symmetry. Then, we assume that the plane of polarization of the incident wave coincides with the plane of symmetry. The simplest example is scattering by a sphere. However, we emphasize that although the specific results presented below refer to such an example, this is done only for the sake of simplicity of calculations. The developed phenomenological theory is valid for a body of any shape with a plane of symmetry.

Following the traditional problem formulation~\cite{Bohren1998}, we choose a coordinate system with the $z$-axis directed along the wave vector of the incident wave $\mathbf{k}$, and the $x$-axis coinciding with the direction of oscillations of the vector $\mathbf{ E}$ of this wave. Then, the vector $\mathbf{H}$ of the incident electromagnetic wave occurs to be parallel to the $y$-axis.

\begin{figure*}
  \centering
  \includegraphics[width=.9\textwidth]{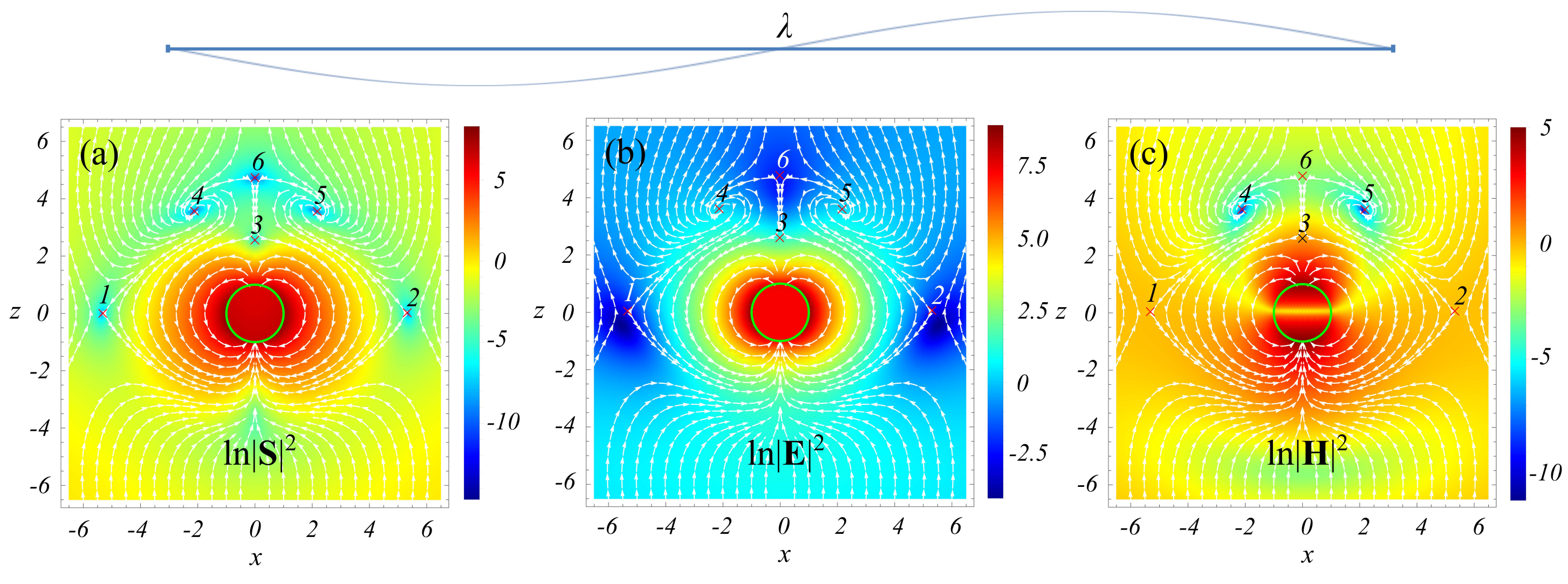}
  \caption{Exact solution of Maxwell's equations. Scattering by a sphere of a plane linearly polarized monochromatic wave in a vacuum. Invariant plane $xz$. Streamlines of the field of the Poynting vector, the values of the logarithm of the square modulus of the Poynting vector (a), electric (b), and magnetic (c) fields (shown in color). The incident wave polarization plane coincides with the one of the figure. The wave vector $\mathbf{k}$ of the incident wave is parallel to the $z$-axis. Size parameter $q=0.3$; $\varepsilon=-2.17$. The solid green line designates the surface of the sphere. The patterns of all fields are symmetric against the plane $x=0$ (perpendicular to the plane of the figure). Crosses (x) mark the position of singular points of the Poynting vector field. Points 4 and 5 are foci; other singularities are saddles; $|\mathbf{S}|^2=0$ at all singular points. The characteristic scale of the field structure is much smaller than the radiation wavelength. The latter's size is shown in the upper part of the figure~\cite{Tribelsky2022}. See text for details.
  }  \label{fig:sphere}
\end{figure*}
   
In the general case, in this problem, the Poynting vector streamlines are essentially three-dimensional. However, it follows from the symmetry that the $xz$ plane is invariant: the streamlines belonging to this plane are two-dimensional. If the scattering body, in addition to the mirror symmetry against the $xz$ plane, is also symmetric with respect to the $yz$ plane (as is the case for a sphere), then this plane is also invariant. Nevertheless, although any restrictions on the positions of the singularities of the $\mathbf{S}(\mathbf{r})$ field are unknown (and there are reasons to expect that such restrictions do not exist), so far, only singularities belonging to an invariant plane parallel to the polarization plane have been observed~\cite{Wang2004,Bashevoy2005,tribelsky2006anomalousPRL,luk2007peculiaritiesJOA,luk2007peculiaritiesCOLA,Tribel:2022_UFN,Tribelsky2022,Tribelsky2022_Nanomat_diss}. Therefore, just these singularities will be discussed below.

Choosing the coordinate system origin at a singular point, taking into account that, due to the invariance of the $xz$ plane, the component $S_y(x,0,z)\equiv 0$, and expanding the components of the vector $\mathbf{S}$ into a series in small departures from the singularity, we conclude that in this case the equation \eqref{eq:stream_vector} takes the form:
\begin{eqnarray}
  \frac{dx}{dt} &=& S_x(x,y,z) \approx s_x^{(x)}x + s_{x}^{(y)}y + s_{x}^{(z)}z, \label{eq:dx/dt} \\
  \frac{dy}{dt} &=& S_y(x,y,z) \approx s_{y}^{(y)}y, \label{eq:dy/dt}\\
  \frac{dz}{dt} &=& S_z(x,y,z) \approx s_{z}^{(x)}x + s_{z}^{(y)}y + s_{z}^{(z)}z, \label{eq:dz/dt}
\end{eqnarray}
where $s_{x_n}^{(x_m)} \equiv \left(\frac{\partial S_{x_n}}{\partial x_m}\right)_{\! s}$. Here index {\it s} means that we calculate the derivative at the singular point, and $x_m$ designates any of the three components of vector {\bf r}.

In the standard way, looking for the solution of the system of equations \eqref{eq:dx/dt}--\eqref{eq:dz/dt} in the form $x_n=x_{n0}\exp(\kappa t)$, \mbox{$x_ {n0}=const_n$} and equating the determinants of the resulting system of algebraic equations to zero, we obtain the roots of the characteristic equation
\begin{eqnarray}
 \!\!\!\!\!\!& & \kappa_{1,2}=\gamma \pm \alpha,\;\; \kappa_3 = s_{y}^{(y)}, \label{eq:kappa_123} \\
 \!\!\!\!\!\!& & \gamma = \frac{s_{x}^{(x)}+s_{z}^{(z)}}{2},\; \alpha = \frac{\sqrt{\left(s_x^{(x)}-s_z^{(z)}\right)^2+4s_x^{(z)}s_z^{(x)}}}{2}. \label{eq:alpha_gamma_3D}
\end{eqnarray}
Note that $\kappa_3$ is always a real quantity, which in the vicinity of the considered singularity corresponds to the exponential repulsion of streamlines from the invariant plane at \mbox{$\kappa_3>0$} and their exponential approach to it at $\kappa_3<0$.

Let us apply the energy conservation law, according to which ${\rm div}\;\!\mathbf{S}=-Q$, where $Q$ is the density of the dissipated electromagnetic energy. Since we are now discussing the non-dissipative case, $Q=0$. As for ${\rm div}\;\!\mathbf{S}$, in the considered approximation, \mbox{${\rm div}\;\!\mathbf{S} \approx s_{x}^{ (x)} + s_{y}^{(y)} + s_{z}^{(z)}$.} In this case, it follows from expressions \eqref{eq:kappa_123}--\eqref{eq:alpha_gamma_3D} that $\kappa_3$ and $\gamma$ have opposite signs. In other words, if, in the invariant plane, the energy flow is directed toward the singularity, then the streamlines go away from the singular point in the perpendicular direction and vice versa. This property is consistent with the integral form of the law of conservation of energy, which states that for a stationary problem in a non-dissipative medium, the energy flux through any closed surface (including a surface surrounding the singularity) must vanish. 

Now, we can discuss the types of singularities in the invariant plane. As an example, Fig.~\ref{fig:sphere} shows the field pattern that occurs in this case for $\varepsilon = -2.17$ and the size parameter $q\equiv kR=0.3$, where $k = \omega/ c$ stands for the wavenumber of the incident wave in a vacuum and $\omega$ is the circular frequency of the incident wave.

Fig.~\ref{fig:sphere}a, exhibits six singular points belonging to the $xz$ invariant plane. Two (numbered 4 and 5) are stable foci; the others are saddles. We emphasize that it is clear from the discussion above that in three-dimensional space (3D), stable in the invariant plane, foci 4 and 5 are unstable in the direction perpendicular to it, i.e., they are complex saddle-focus singular points.

The calculations show that the Poynting vector vanishes at all singular points, as it should. To understand which electromagnetic type (field-induced or polarization-induced) these singularities belong to, we superimpose the streamlines of Fig.~\ref{fig:sphere}a with the profile of the fields $|\mathbf{E}|^2$ and $| \mathbf{H}|^2$, see Fig.~\ref{fig:sphere}b,c. It can be seen that the foci are $H$-induced singularities, and the saddles are polarization-induced.

To interpret this result, we note that polarization-induced singularities correspond to the local formation of standing waves when pairs of waves of the same amplitude pass the singularity in opposite directions. It implies that the singularity has singled out directions. A saddle has such directions: these are the ``whiskers'' (stable and unstable manifolds) of the separatrices. Unlike saddles, foci has no preferred directions, which makes it difficult for standing waves to form near a focus~\footnote{An important comment on this argument is given in Conclusions.}. Note that such a distinction between foci and saddles is topologically stable, since these singularities are not topologically equivalent, i.e., a transformation of the coordinate system cannot reduce them to each other~\cite{ ODE:Arnold}. 

We particularly emphasize that the discussed area of standing wave formation has essentially sub-wavelength size, which can be clearly seen in Fig.~\ref{fig:sphere} where, for clarity, the wavelength of the incident radiation is shown at the same scale as that for the field structures.

An essential feature of the results discussed is that the field-induced singularities (foci) are associated with the vanishing of that field, which is {\it perpendicular\/} to the invariant plane. It is $\mathbf{H}$ in the above case. However, such a structure of field-induced singularities is generic for singularities belonging to invariant planes. If the polarization and invariant planes are perpendicular, the singularities occur $E$-induced. To verify this fact, we consider the scattering of light by a cylinder.

\subsection{Cylinder\label{sec:cyl}}  

As in the previous section, the developed phenomenological theory is valid for an arbitrary cross-section-shaped infinite cylinder. The employed examples for a right circular cylinder are given only to simplify the calculations. Let us start with the cases of perpendicular incidence and two independent polarizations of the incident wave: TE and TM, see Fig.~\ref{fig:TE_TM}, where the orientation of the coordinate axes and the $\mathbf{k}$-vector corresponds to the generally accepted one, see, e.g.~\cite{Bohren1998}. 

In this case, the dependence of the fields on the $z$-coordinate vanishes, the problem becomes two-dimensional (2D), the expansions for the Poynting vector components near the singularity takes the form $S_x(x,y) \approx s_x^{(x)}x + s_{x }^{(y)}y;$ \mbox{$S_y(x,y) \approx s_{y}^{(x)}x + s_{y}^{(y)}y$}, and the condition ${\rm div}\;\!\mathbf{S}=0$ reduces to $s_x^{(x)}+s_y^{(y)}=0$. As a result, for the roots of the characteristic equation, instead of \eqref{eq:kappa_123},~\eqref{eq:alpha_gamma_3D}, we get the following expression:
 
\begin{figure}
\includegraphics[width=.95\columnwidth]{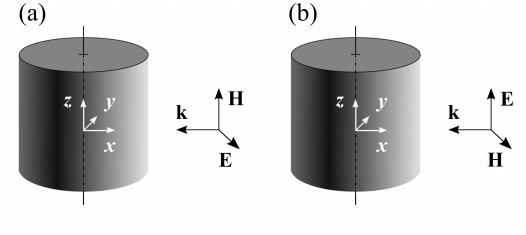}
\centering
\caption{Mutual orientation of the cylinder, coordinate axes, and vectors {\bf k}, {\bf E}, {\bf H} of the incident wave. (a) TE polarization, \mbox{ (b) TM polarization~\cite{Tribelsky2022,Bohren1998}.}\label{fig:TE_TM}}
\end{figure}

\begin{figure*}
\includegraphics[width=.95\textwidth]{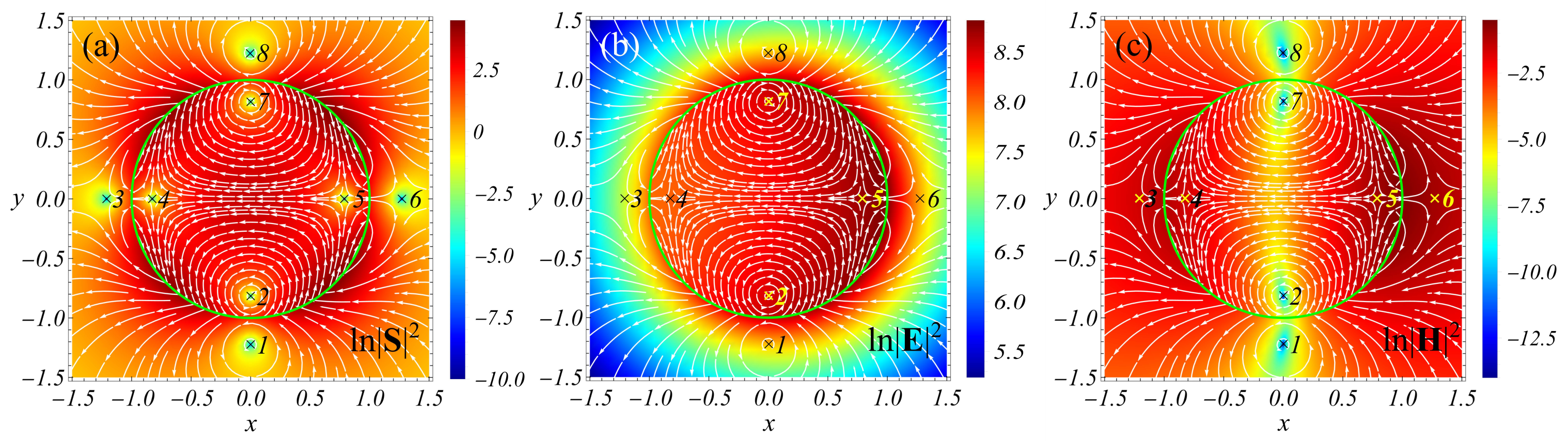}
\caption{Scattering of a plane TE-polarized monochromatic electromagnetic wave by a right circular cylinder. Streamlines of the Poynting vector, as well as fields $|\mathbf{S}|^2$, $|\mathbf{E}|^2$, $|\mathbf{H}|^2$ on a logarithmic scale; $q=0.1$; $\varepsilon=-1$. The incident plane wave propagates in the negative direction of the $x$ axis; see Fig.~\ref{fig:TE_TM}. The field structures are mirror symmetric against the plane $y=0$ perpendicular to the plane of the figure. The crosses (x) denote the singularities of the field $\mathbf{S}$. A green circle marks the cylinder surface. The centers (points 1,2,7,8) correspond to $H$-induced singularities. Saddles (points 3--6) are polarization-induced~\cite{Tribelsky2022}.}\label{fig:Cylinder_TE}
\end{figure*}

\begin{figure*}
\includegraphics[width=.95\textwidth]{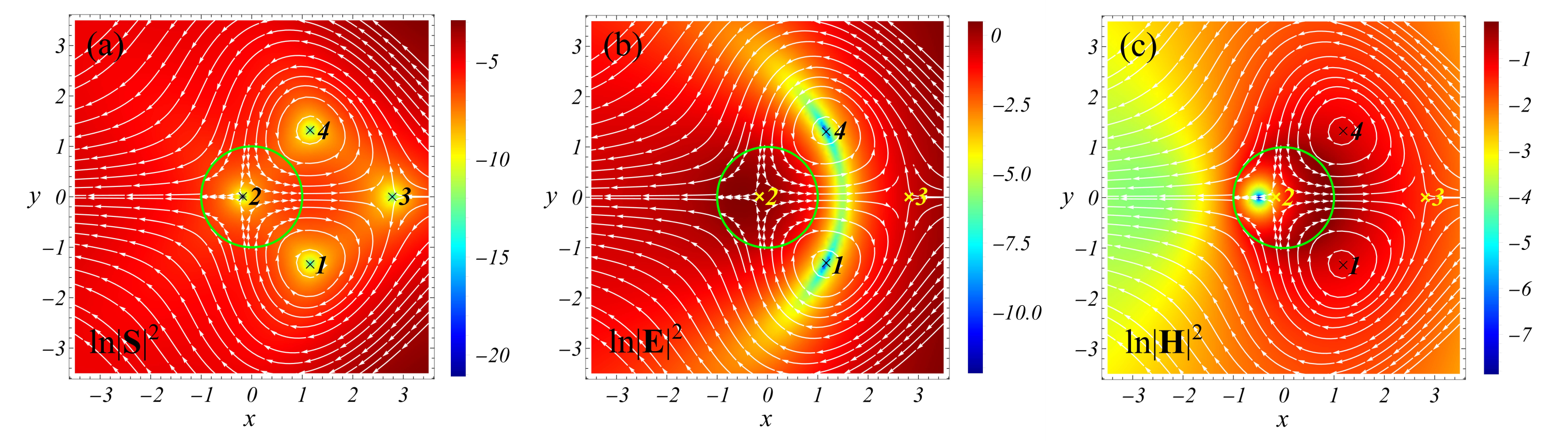}
\caption{The same as that in Fig.~\ref{fig:Cylinder_TE} with TM polarization of incident radiation and $q=0.3$, $\varepsilon=16$. In this case, the centers (points 1, 4) correspond to $E$-induced singularities~\cite{Tribelsky2022}.}\label{fig:Cylinder_TM}
\end{figure*}     

\begin{figure*}
  \centering
  \includegraphics[width=\textwidth]{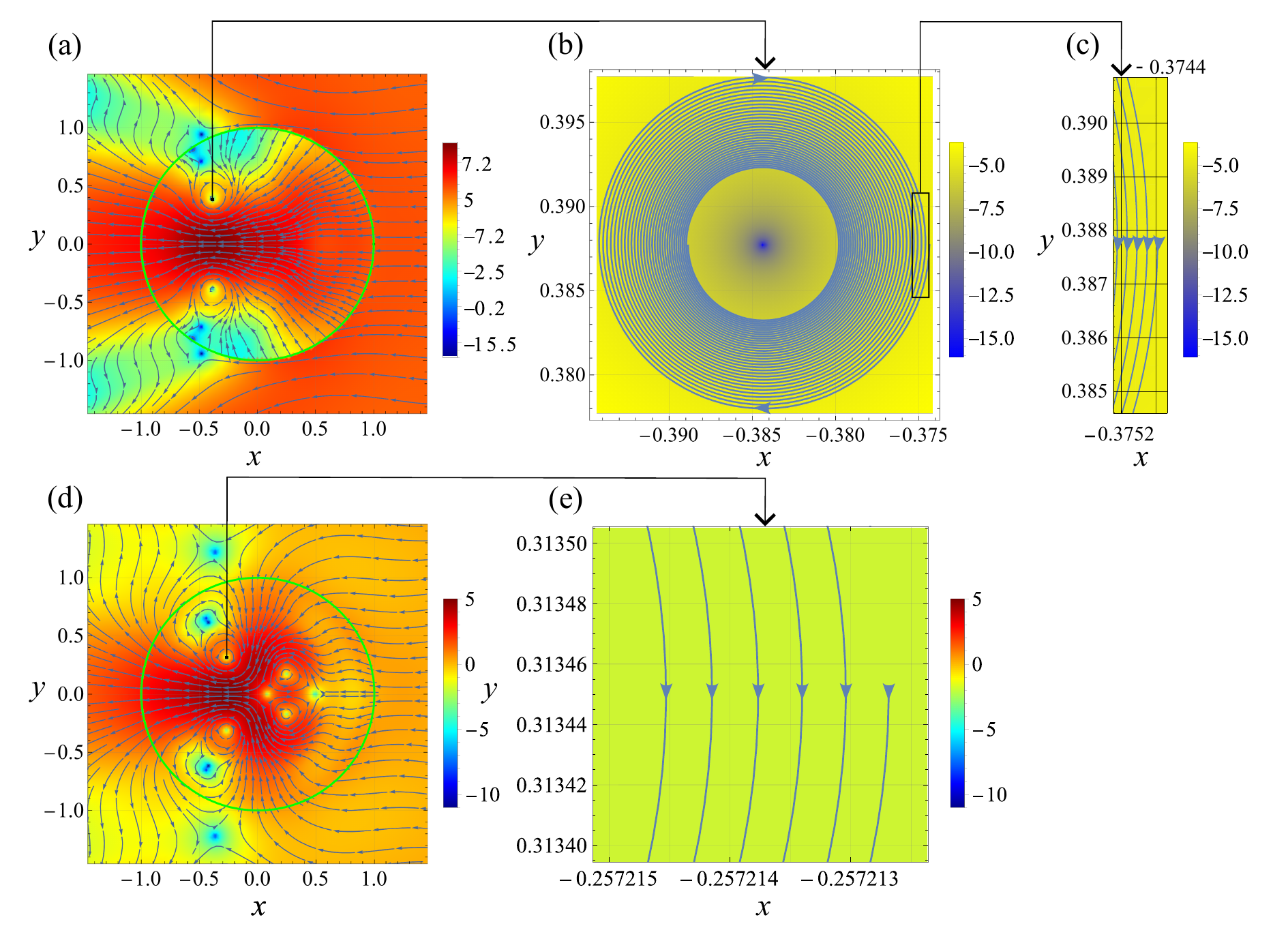}
  \caption{The Poynting vector field and its streamlines at light scattering by a germanium cylinder in a vacuum. The plane linearly polarized monochromatic incident wave propagates in the negative direction of the $x$-axis.  A green circle designates the cylinder surface. The wavelength is $\lambda=\lambda_1=1590$ nm, and the complex permittivity $\varepsilon(\lambda_1)\approx 17.775+i0.024$. The size parameter $q=1.62$. The field pattern exhibits symmetry against the plane $y=0$, which is perpendicular to the plane of the figure. Panels (a)-(c) correspond to TE polarization of the incident radiation, while panels (d) and (e) represent TM polarization. Panel (b) offers a closer view of the singularity region, marked in panel (a) with a small black rectangle. The same applies to panels (d) and (e). Panel (c) provides a zoomed-in view of the region marked in panel (b) with a rectangle. See text for details. Note the significant difference in scales between panels (c) and (e)~\cite{Tribelsky2022_Nanomat_diss}.}\label{fig:1590}
\end{figure*}

\begin{figure*}
  \centering
  \includegraphics[width=\textwidth]{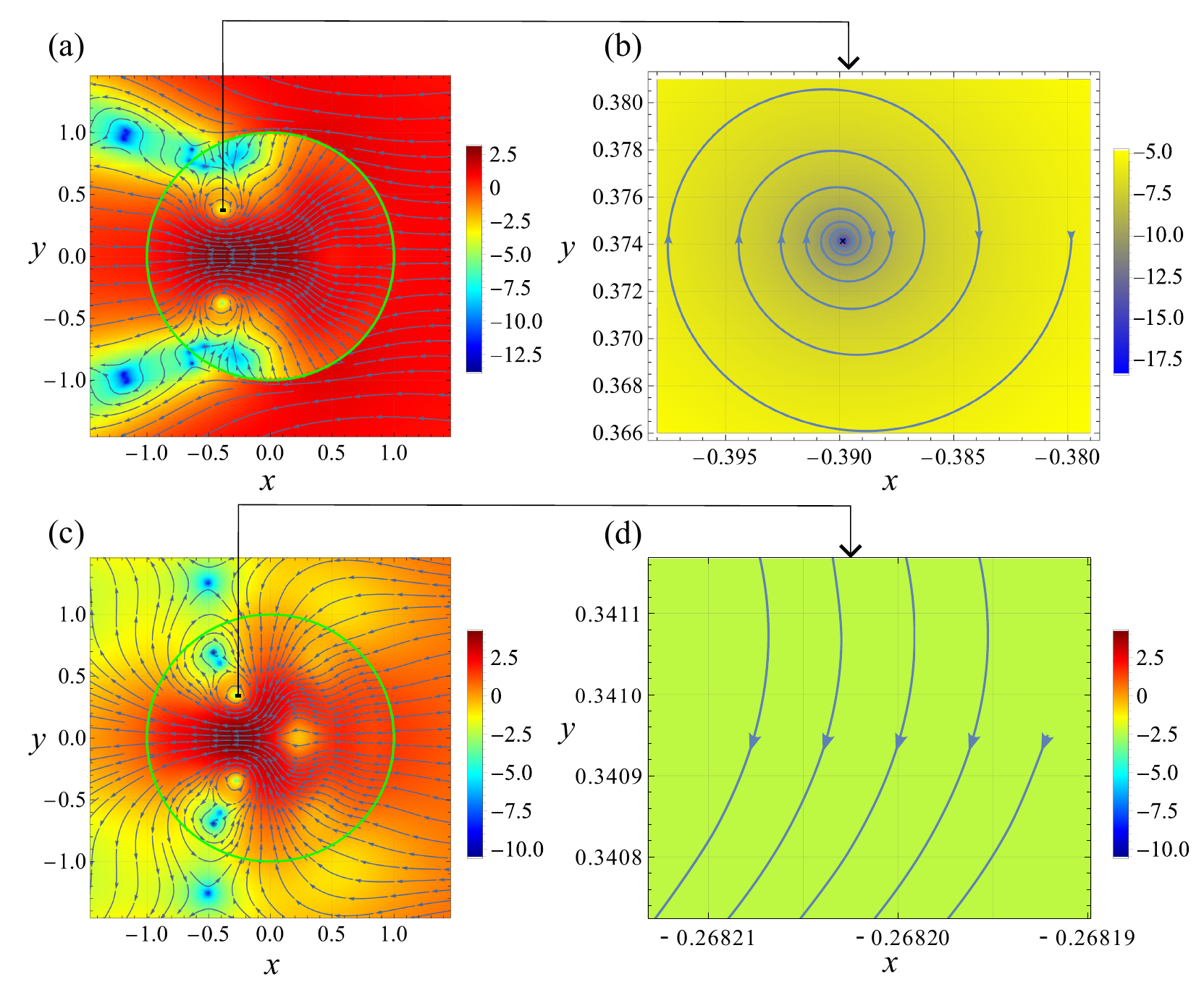}
  \caption{The same as that in Fig~\ref{fig:1590}a,b,d,e, at $\lambda=\lambda_2=1494$ nm; \mbox{$\varepsilon(\lambda_2)\approx 17.983+i0.483$}~\cite{Tribelsky2022_Nanomat_diss}.}\label{fig:1494}
\end{figure*}

\begin{figure}
  \centering
  \includegraphics[width=.5\columnwidth]{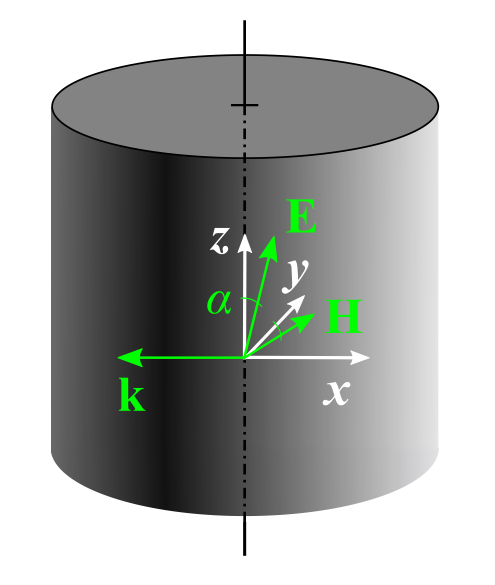}
  \caption{Mutual orientation of the cylinder, coordinate frame, and vectors $\mathbf{k}$, $\mathbf{E}$, $\mathbf{H}$ of the incident electromagnetic wave at an arbitrary orientation of its polarization plane against the cylinder axis~\cite{Tribel:2023_arXiv}.}\label{fig:Orientation}
\end{figure}

\begin{figure*}
  \centering
  \includegraphics[width=\textwidth]{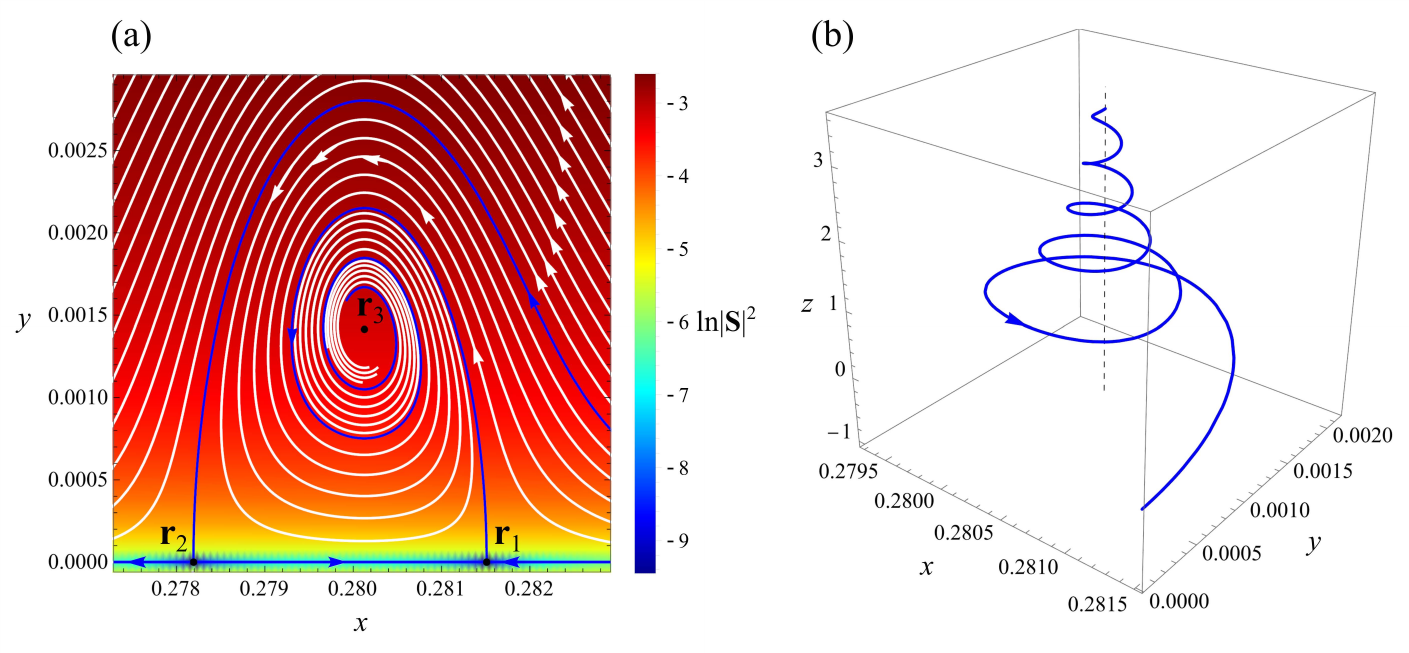}
  \caption{The field of the Poynting vector and its streamlines inside a Germanium cylinder. The wave vector of an incident plane linearly polarized wave is antiparallel to the $x$ axis; see Fig.~\ref{fig:Orientation}. The axis of the cylinder makes angle \mbox{$\alpha = 45.403^\circ$} with the polarization plane; \mbox{$\varepsilon = 17.775+0.024 i$}; \mbox{$q = 1.62$}; (a) 2D projection of the streamlines onto the plane perpendicular to the cylinder axis. The projection has three singularities marked with black dots: two saddles ($\mathbf{r}_{1,2}$) belonging to the $x$-axis and a stable focus outside it ($\mathbf{r}_3$). The latter is a false singularity: it is regular in 3D space. This is clearly seen in panel (b), which shows a 3D image of a part of the separatrix whisker emerging in panel (a) from saddle $\mathbf{r}_{1}$ towards $\mathbf{r}_{3}$.  It is a spiral asymptotically going to infinity, which winds on a straight line parallel to the cylinder axis. The projection of this line onto the $xy$ plane gives the point $\mathbf{r}_{3}$~\cite{Tribel:2023_arXiv}. See text for details. Note the significantly different scales of the $x$-, $y$-, and $z$-axis in panel (b). }\label{fig:2D-3D}
\end{figure*}

\begin{equation}\label{eq:kappa12_cyl}
  \kappa_{1,2} = \pm \sqrt{\left(s_x^{(x)}\right)^2+s_x^{(y)} s_y^{(x)}}
\end{equation}
Thus, in the 2D case, the law of conservation of energy leads to a significant reduction in the possible types of singularities, namely: for \mbox{$\left(s_x^{(x)}\right)^2+s_x^{(y)} s_y^{(x)} > 0$} they are saddles, and for \mbox{$\left(s_x^{(x)}\right)^2+s_x^{(y)} s_y^{(x)} < 0$} the singularities are centers. The existence of other singular points is forbidden. The case $\left(s_x^{(x)}\right)^2+s_x^{(y)} s_y^{(x)}=0$ is degenerate. To analyze this case, higher-order-terms in the \mbox{$S_{x,y}(x,y)$-expansion} must be considered.

As an example, Fig.~\ref{fig:Cylinder_TE}, \ref{fig:Cylinder_TM} show the structures of the fields and streamlines of the Poynting vector in the scattering of TE and TM polarized radiation by a cylinder. These examples are in complete agreement with the general considerations above.

\section{Dissipative effects\label{sec:diss}}

As has been said above, the medium embedding the scattering particle must be non-dissipative. Thus, only the particle itself can have dissipative properties. Therefore, the singularities discussed in this section must be situated inside such a particle. How does dissipation affect the properties of these singularities? 

First, note that the streamlines cannot form a family of closed-loop lines since dissipative losses now accompany the motion along them. Because of that, all center-type singularities turn into foci.~\footnote{This is entirely true in 2D. In 3D, the field $\mathbf{S}(\mathbf{r})$, in principle, may have such a pattern that the energy inflow along the directions transverse to this plane exactly compensates the dissipative losses for trajectories in the invariant plane. In this exceptional situation, the existence of closed-loop trajectories becomes possible.}

In the quantitative description of the problem under consideration, the only difference between dissipative and nondissipative media is the formulation of the energy conservation law, which in a dissipative medium is of the form ${\rm div}\;\!\mathbf{S}=-Q < 0$. For $Q$, in the chosen dimensionless variables, the following formula holds: $Q=\varepsilon''q|\mathbf{E}|^2$ (remember: $q = kR$ is the size parameter).

Note that the dissipation is related only to the electric field since the magnetic permeability at optical frequencies equals unity. This leads to asymmetry between the $\mathbf{E}$ and $\mathbf{H}$ fields. In particular, the effect of dissipation on $H$-induced singularities is expected to be much larger than that on $E$-induced singularities.

Indeed, in the equation ${\rm div}\;\!\mathbf{S}(\mathbf{r})=-Q(\mathbf{r})$, it is necessary for the right-hand and left-hand sides to be of the same order of magnitude. In the approximation being considered, ${\rm div}\;\!\mathbf{S} \approx \sum_{n}s_{x_n}^{(x_n)} = const$. Consequently, the right-hand side of the equation should be represented as $Q(\mathbf{r}) \approx Q(\mathbf{r}_s) = const$, where $\mathbf{r}_s$ denotes the coordinates of the singular point. Hereafter, $Q$ will specifically denote $Q(\mathbf{r}_s)$. In $H$- and polarization-induced singularities, the electric field does not turn to zero; thus, $Q\neq 0$. Further analysis for a sphere and cylinder is convenient to conduct separately.

\subsection{Sphere \label{sec:sphere_diss}}  

For a sphere, the characteristic equation roots are still given by Eqs.~\eqref{eq:kappa_123},~\eqref{eq:alpha_gamma_3D}. However, now instead of the expression $\sum_{n}^{3}s_{x_n}^{(x_n)} =0$, the condition $\sum_{n}^{3}s_{x_n}^{(x_n)}=-Q$ should be in use. This additional condition imposes one constraint on the three independent coefficients $s_{x_n}^{(x_n)}$. Therefore, the problem still has enough ``degrees of freedom,'' so that this constraint does not change its qualitatively. 
Specifically, in expressions \eqref{eq:kappa_123},~\eqref{eq:alpha_gamma_3D}, the parameter $\gamma$ can have both positive and negative signs, indicating that streamlines can either approach or move away from the singularity. Escaping trajectories do not contradict the presence of dissipative losses since energy inflow occurring along directions transversal to the invariant plane compensates for these losses.

It is important to note that the vanishing of $\gamma$ at $s_{x}^{(x)}+s_{z}^{(z)}=0$ does not imply that the streamlines are closed-loop, as higher-order terms, not considered in the linear approximation, affect their behavior. A more detailed discussion on this matter is presented below; see Sec.~\ref{sec:cyl_diss}.

Additionally, note that the integral form of the energy conservation law makes it possible to derive a simple formula relating the various components of the Poynting vector near the singularity. To obtain it, we consider an imaginary right circular cylinder surrounding the singularity. Its bases with a small radius $r$ are situated symmetrically against the invariant plane at distances $\pm z$ ($z\sim r$), while its axis passes through the given singular point perpendicular to the invariant plane. We introduce a local cylindrical coordinate system where the center coincides with the singularity, and the axis is aligned along the axis of the imaginary cylinder. Taking into account that the Poynting vector flux through the entire surface of the cylinder is equal to the power dissipated within its volume, we readily obtain
\begin{equation}\label{eq:<Sr>}
  \langle S_r\rangle \approx - \frac{rs_y^{(y)}+Q}{2},
\end{equation}  
where $\langle S_r\rangle$ is the angular-averaged radial component of the Poynting vector in the local coordinate system. Expression \eqref{eq:<Sr>} is valid for any type of singular point belonging to the invariant plane.

\subsection{Cylinder \label{sec:cyl_diss}} 

In contrast to a sphere, the impact of dissipation is much more pronounced in the case of scattering by an infinite cylinder. It happens owing to the reduction of the spatial dimension from 3D to 2D. We begin the consideration with the case of the incident radiation's pure TE or TM polarization. Firstly, it should be noted that the expressions for the roots of the characteristic equation are now analogous to the expressions for $\kappa_{1,2}$  for the sphere; see Eq.~\eqref{eq:kappa_123}. In this case, $\gamma = -Q/2$~\cite{Tribelsky2022_Nanomat_diss}. The value of $\alpha$ can be either purely real or purely imaginary. For $Q>0$ and a TE-polarized incident wave, where all singular points are $H$-induced, only saddles, stable foci, and nodes can occur as singularities. These quantitative results agree perfectly with the qualitative arguments mentioned earlier.

For TM polarization of the incident radiation, the electric field at a singular point turns to zero (\mbox{$Q=0$}). The case looks equivalent to the non-dissipative limit. However, it is not. The critical distinction is that the dissipation is zero only at the singularity itself. In its vicinity, the amplitude of the electric field is small but not equal to zero. It is described by higher-order terms dropped in the discussed linear theory. Accounting for these terms results in closed-loop streamlines (singularity of the center type) transforming into spiral-shaped ones that converge towards the singular point, similar to the TE polarization case. The difference, however, is that the dissipation is extremely weak now. It gives rise to a much smaller pitch for these spirals than that at the same $\varepsilon$ for TE polarization. 

As an example, Figs.~\ref{fig:1590},~\ref{fig:1494} depict the fields and streamlines calculated for a Germanium cylinder at wavelengths $\lambda_1 = 1590$ nm and $\lambda_2 = 1494$ nm, respectively.  At the calculations, we employ the actual permittivity values for these wavelengths $\lambda$~\cite{Polyanskiy}. They are as follows: \mbox{$\varepsilon(\lambda_1)\approx 17.775+i0.024$}, while \mbox{$\varepsilon(\lambda_2)\approx 17.983+i0.483$}. In other words, the real parts of $\varepsilon(\lambda_{1,2})$ are close to each other, while their imaginary parts differ by more than twenty times. Such a choice facilitates an exploration of the dissipation effects at different values of the dissipative constant while keeping other parameters of the problem practically fixed.

The size parameter $q$ for both values of $\lambda$ is the same and equal to 1.62, which, for the given values of $\varepsilon(\lambda_{1,2})$, corresponds to the vicinity of the dipole resonance for both TE and TM polarizations. This fact ensured the similarity of the $\mathbf{S}(\mathbf{r})$ field patterns for both polarizations. Note also that $R/\lambda_{1,2} \approx 0.26$, i.e., for the incident radiation, such a cylinder is essentially a sub-wavelength particle. Although the singularities marked in Figs.~\ref{fig:1590}a,~\ref{fig:1590}d, \ref{fig:1494}a, and \ref{fig:1494}c look like centers, zoom shows that, in fact, they are stable foci. In agreement with the above discussion, the pitch of the helical streamlines in the case of TE polarization turns out to be much larger than in the case of TM. In both cases, the pitch increases with increasing value of the imaginary part of the permittivity, cf. Fig.~\ref{fig:1590} and Fig.~\ref{fig:1494}.

\section{Symmetry breaking effects \label{sec:oblique}} 
So far, we have considered highly symmetric solutions. However, in an actual experiment, symmetry is always violated. In this context, the topological stability of the results under symmetry breaking is essential. The key question here is: How universal are the results discussed above, and how do the properties of the singularities change when the symmetry is broken? To elucidate this issue, we have to distinguish between weak symmetry breaking due to a non-ideal shape of the laser beam and/or the scattering particle, fluctuations in permittivity, etc., and substantial violations, for example, when a particle of an arbitrary shape scatters light.

As for substantial symmetry breaking, although the author is unaware of any reliable example of singularities of the Poynting vector field in such problems, there is no reason why such singularities could not occur. If such a singularity does occur, it must be essentially three-dimensional because of the lack of symmetry. In general, the Jacobian $J\equiv\left(\frac{\partial(S_x,S_y,S_z)}{\partial(x,y,z)}\right)_s$ has nine real nonzero entries. The corresponding cubic characteristic equation has three roots $\kappa_{1,2,3}$, of which either all three are real, or one is real, and two are complex conjugate. In this case, the single condition $Sp\{J\}=-Q$ imposed on the Jacobian entries does not lead to restrictions on the signs of the roots of the characteristic equation. The situation is the same as that described above in the case of a sphere. If three real roots have the same sign, it is a node (stable for $\kappa_{1,2,3} < 0$ and unstable for the opposite sign). If one of the roots has a sign opposite to the other two, this is a saddle-node. Finally, if two roots are complex, the singularity is a saddle-focus. This is all we can say about a general case singularity at substantial symmetry breaking. 

The case of weak symmetry breaking is much more interesting. This issue is inspected in paper~\cite{Tribel:2023_arXiv}. The difference in the problem formulation in this study with the ones discussed above is that while in Ref.~~\cite{Tribel:2023_arXiv} the wave vector $\mathbf{k}$ still remains perpendicular to the cylinder axis, the latter made an arbitrary angle $\alpha$ with the polarization plane; see Fig.~\ref{fig:Orientation}. This case is one of the simplest versions of the problem with several symmetry groups, some of which are violated in a controlled manner while others remain non-broken. Specifically, in the example under consideration, one can control the violation of mirror symmetry against the $xy$ plane by changing the angle $\alpha$ while preserving the symmetry against arbitrary translations along the cylinder axis.

Because of the mentioned translational symmetry, the fields $\mathbf{E},\;\mathbf{H}$, and $\mathbf{S}$ can depend only on $x$ and $y$ but not on $z$. Since all three components of the Poynting vector vanish at a singular point, its coordinates $x,y$ must satisfy {\it three\/} independent equations: $S_x(x,y)=S_y(x,y)=S_z(x,y)=0$. However,  {\it two\/} variables $x$ and $y$ cannot satisfy three independent equations simultaneously. Such a system is overdetermined and has no solutions.

Seemingly, it leads to the conclusion that the problem has no singularities. However, this is not quite the case. The point is that one or more components of the Poynting vector can identically vanish because of the remaining problem symmetry. It reduces the number of equations in the system, which determine the position of the singularity, making them compatible.

A detailed analysis of the symmetry of this problem for a right circular cylinder, taking into account the restrictions imposed by the boundary conditions on its surface, shows that the components of the Poynting vector satisfy the following relations:~\cite{Tribel:2023_arXiv} 
\begin{eqnarray}
  & & (S_x(x,y),S_y(x,y),S_z(x,y)) = \nonumber\\
  & & (S_x(x,-y),-S_y(x,-y),-S_z(x,-y)). \label{eq:S_y->-y}
\end{eqnarray}

These relations, in particular, show that $S_z\equiv 0$ at $y=0$. Then, singular points may appear on the $x$-axis. Note that such singularities can only be nodes and saddles since the topological structure of centers and foci does not satisfy the symmetry of the $S_x$ and $S_y$ components against the $y \rightarrow -y$ transformation.

However, now, along with true singularities, ``false'' singularities appear. These points are singular in the 2D projection of streamlines onto the $xy$ plane but have a non-zero $z$-component of the Poynting vector, so in 3D, they are regular points, the set of which creates a vertical line parallel to the $z$ axis. Figure~\ref{fig:2D-3D} shows an example of such a false singularity and two true singular points belonging to the $x$-axis.

Note that similar ``false'' singularities are well known and often encountered, for example, in the theory of paraxial beams.  These points are singular only in the projection of streamlines onto a plane perpendicular to the beam axis. In contrast, the component of the Poynting vector along the beam axis remains finite, i.e., in 3D, such ``singularities'' are regular points; see, for example, publications~\cite{bekshaev2007transverse,bekshaev2011internal}.

The essential difference between such ``singularities'' and those discussed in this section is that for the former the beam symmetry imposes the direction in which the Poynting vector component does not vanish. This singled-out direction is the beam axis: the non-vanishing component of the Poynting vector either coincides with the average direction of propagation of the incident radiation or, in some extraordinary situations, it has the opposite direction. In the cases discussed here, this restriction does not exist. In particular, the nonzero component of the Poynting vector at the point $\mathbf{r}_3$ in Fig.~\ref{fig:2D-3D}a is {\it perpendicular\/} to the wave vector of the incident plane wave, oriented antiparallel to the $x$-axis. It is clearly seen in Fig.~\ref{fig:2D-3D}b.

Fig.~\ref{fig:2D-3D}b also shows that the depicted three-dimensional streamline has no translational symmetry along the \mbox{$z$-axis.} It is easy to understand. Indeed, in the general case, at a regular point, all three components of the vector $\mathbf{S}$ have different nonzero values. Then, as it follows from Eq.~\eqref{eq:stream_vector}, for the streamline emerging from this regular point, the dependence on ``time'' $t$ for each of the three spatial coordinates is individual. This results in a substantially three-dimensional shape of the given line. On the other hand, since the projection of this line onto the $xy$ plane corresponds to a converging spiral, see Fig.~\ref{fig:2D-3D}a, such a line cannot be invariant against translations along the \mbox{$z$-axis.} From these considerations, it is evident that three-dimensionality and the absence of translational symmetry are typical properties of streamlines in such a problem, except straight lines parallel to the $x$-axis lying in the $xz$ plane, which, due to the conditions~\eqref{eq:S_y->-y} are also exact solutions of Eq.~\eqref{eq:stream_vector}.

The conclusion about the absence of translational symmetry for three-dimensional streamlines seems contradictory to the above-noted invariance of the $\mathbf{S}$ field against arbitrary translations along the $z$-axis. Actually, there is no contradiction. The replacement $z \rightarrow z + const$ transforms the streamline into {\it another\/} streamline, the shape of which is identical to the original one. 

These results show that the behavior of streamlines in the vicinity of the singularities discussed here is topologically stable against weak symmetry breaking. Although such symmetry violation leads to the regularization of singular points due to the appearance of a small non-zero component of the Poynting vector transversal to the original invariant plane, the streamline pattern projection onto the original invariant plane does not change not only qualitatively (cf. Fig.~\ref{fig:1590}d and \ref{fig:2D-3D}a), but the quantitative changes remain small as long as the symmetry breaking is small. For more detail, see Ref.~\cite{Tribel:2023_arXiv}.

The constructive use of the problem symmetry also makes it possible to develop a phenomenological theory explaining the sequence of bifurcations leading to the emergence (annihilation) of both false and true singularities when the varying bifurcation parameter is either the angle $\alpha$ or any other parameter of the problem (i.e., $ q$ or $\varepsilon$). However, the author believes that these results should be of interest only to experts in this specific field. Therefore, they will not be discussed here. Readers interested in these issues may find their discussion in Ref.~\cite{Tribel:2023_arXiv}. 

\section{Conclusions \label{sec:concl}}

In conclusion, we note that since polarization-induced singularities are associated with the formation of standing waves in a small neighborhood of a singular point, this implies the existence of singled-out directions in this region, along which pairs of waves propagate in opposite directions. Further reasoning is essentially different in 2D and 3D cases.

In 2D, there are no such directions for center and focus. Therefore, these singularities belong to the field-induced type. As for nodes and saddles, there are no restrictions here, and both types of electromagnetic singularities are possible for them. However, the fields $\mathbf{E}$ and $\mathbf{H}$ enter the problem symmetrically. The vanishing of one of them at a singularity distinguishes that field relative to the other and breaks the $\mathbf{E}-\mathbf{H}$ symmetry. Then, one should expect that if the vanishing does not follow from the topological properties of the singularity (as it is in the case of a center and focus), nodes and saddles should be polarization-induced in the general case. Field-induced singularities for nodes and saddles are exceptional cases associated with a certain degeneracy. One way or another, all the saddles and nodes that we have observed so far in various cases are polarization-induced.

As for 3D, the cubic characteristic equation always has at least one purely real root. Then, in the vicinity of such a singularity, there is a singled-out direction of the eigenvector corresponding to this purely real root. Along this direction, on opposite sides of the singularity, the energy flows are aligned opposite each other. For this reason, the topological structure of the singularity does not impose any restrictions on its electrodynamic type. Nevertheless, in the case of scattering by a sphere, all foci belonging to the invariant plane turned out to be of the field-induced type. It may be due to the symmetry violation between $\mathbf{E}$ and $\mathbf{H}$, introduced by the coincidence of the invariant plane with the plane of polarization.

Note also the problem of the topological charge calculation for these singularities and the issue of its conservation at bifurcations discussed in Ref.~\cite{Tribel:2023_arXiv}. Although there are no fundamental difficulties in finding answers to these questions, at the moment, they remain open and can be considered as issues for future study. 

In conclusion, we again emphasize that the singularities discussed here are not related to any peculiarities of the incident laser beam. They appear spontaneously in the scattering of a plane linearly polarized wave with no singularities and have essentially sub-wavelength characteristic scales.

Thus, the paper reveals the deep connection between the topological structure of the singularities of the Poynting vector field, their electromagnetic type, the law of energy conservation, the symmetry of the problem, and the dimension of space. \pagebreak These results shed new light on the problem of electromagnetic energy circulation in fields of complex configurations and create a basis for their practical implementations in various applications, particularly in tailoring energy flows of a given sub-wavelength structure, which is essential for many nanotechnologies.

The author is grateful to B. Ya. Rubinshtein, for the discussion of this article and valuable comments. This work was financially supported by the Russian Science Foundation under project no. 21-12-00151 (analytical research) and by the Ministry of Science and Higher Education of the Russian Federation under project no. 075-15-2022-1150 (computer calculations and computer graphics). The influence of symmetry effects on the properties of singular points of the Poynting vector was studied with the support of the Russian Science Foundation grant No. 23-72-00037.

\bibliography{Singul_F_2} 

\begin{thebibliography}{44}%
\makeatletter
\providecommand \@ifxundefined [1]{%
 \@ifx{#1\undefined}
}%
\providecommand \@ifnum [1]{%
 \ifnum #1\expandafter \@firstoftwo
 \else \expandafter \@secondoftwo
 \fi
}%
\providecommand \@ifx [1]{%
 \ifx #1\expandafter \@firstoftwo
 \else \expandafter \@secondoftwo
 \fi
}%
\providecommand \natexlab [1]{#1}%
\providecommand \enquote  [1]{``#1''}%
\providecommand \bibnamefont  [1]{#1}%
\providecommand \bibfnamefont [1]{#1}%
\providecommand \citenamefont [1]{#1}%
\providecommand \href@noop [0]{\@secondoftwo}%
\providecommand \href [0]{\begingroup \@sanitize@url \@href}%
\providecommand \@href[1]{\@@startlink{#1}\@@href}%
\providecommand \@@href[1]{\endgroup#1\@@endlink}%
\providecommand \@sanitize@url [0]{\catcode `\\12\catcode `\$12\catcode
  `\&12\catcode `\#12\catcode `\^12\catcode `\_12\catcode `\%12\relax}%
\providecommand \@@startlink[1]{}%
\providecommand \@@endlink[0]{}%
\providecommand \url  [0]{\begingroup\@sanitize@url \@url }%
\providecommand \@url [1]{\endgroup\@href {#1}{\urlprefix }}%
\providecommand \urlprefix  [0]{URL }%
\providecommand \Eprint [0]{\href }%
\providecommand \doibase [0]{https://doi.org/}%
\providecommand \selectlanguage [0]{\@gobble}%
\providecommand \bibinfo  [0]{\@secondoftwo}%
\providecommand \bibfield  [0]{\@secondoftwo}%
\providecommand \translation [1]{[#1]}%
\providecommand \BibitemOpen [0]{}%
\providecommand \bibitemStop [0]{}%
\providecommand \bibitemNoStop [0]{.\EOS\space}%
\providecommand \EOS [0]{\spacefactor3000\relax}%
\providecommand \BibitemShut  [1]{\csname bibitem#1\endcsname}%
\let\auto@bib@innerbib\@empty
\bibitem [{\citenamefont {Mokhun}\ \emph {et~al.}(2007)\citenamefont {Mokhun},
  \citenamefont {Khrobatin}, \citenamefont {Mokhun},\ and\ \citenamefont
  {Viktorovskaya}}]{Mokhun2007}%
  \BibitemOpen
  \bibfield  {author} {\bibinfo {author} {\bibfnamefont {I.}~\bibnamefont
  {Mokhun}}, \bibinfo {author} {\bibfnamefont {R.}~\bibnamefont {Khrobatin}},
  \bibinfo {author} {\bibfnamefont {A.}~\bibnamefont {Mokhun}},\ and\ \bibinfo
  {author} {\bibfnamefont {J.}~\bibnamefont {Viktorovskaya}},\ }\href
  {https://www.researchgate.net/profile/Igor-Mokhun/publication/26538197_The_behavior_of_the_Poynting_vector_in_the_area_of_elementary_polarization_singularities/links/00b495226f038411c2000000/The-behavior-of-the-Poynting-vector-in-the-area-of-elementary-polarization-singularities.pdf}
  {\bibfield  {journal} {\bibinfo  {journal} {Optica Applicata}\ }\textbf
  {\bibinfo {volume} {37}},\ \bibinfo {pages} {261} (\bibinfo {year}
  {2007})}\BibitemShut {NoStop}%
\bibitem [{\citenamefont {Novitsky}\ and\ \citenamefont
  {Barkovsky}(2009)}]{Novitsky2009}%
  \BibitemOpen
  \bibfield  {author} {\bibinfo {author} {\bibfnamefont {A.~V.}\ \bibnamefont
  {Novitsky}}\ and\ \bibinfo {author} {\bibfnamefont {L.~M.}\ \bibnamefont
  {Barkovsky}},\ }\href {https://doi.org/10.1103/physreva.79.033821} {\bibfield
   {journal} {\bibinfo  {journal} {Physical Review A}\ }\textbf {\bibinfo
  {volume} {79}},\ \bibinfo {pages} {033821} (\bibinfo {year}
  {2009})}\BibitemShut {NoStop}%
\bibitem [{\citenamefont {Dennis}\ \emph {et~al.}(2009)\citenamefont {Dennis},
  \citenamefont {O'Holleran},\ and\ \citenamefont {Padgett}}]{Dennis2009}%
  \BibitemOpen
  \bibfield  {author} {\bibinfo {author} {\bibfnamefont {M.~R.}\ \bibnamefont
  {Dennis}}, \bibinfo {author} {\bibfnamefont {K.}~\bibnamefont {O'Holleran}},\
  and\ \bibinfo {author} {\bibfnamefont {M.~J.}\ \bibnamefont {Padgett}},\
  }\href {https://doi.org/10.1016/S0079-6638(08)00205-9} {\emph {\bibinfo
  {title} {Progress in Optics; Chapter 5 Singular Optics: Optical Vortices and
  Polarization Singularities}}},\ edited by\ \bibinfo {editor} {\bibfnamefont
  {E.}~\bibnamefont {Wolf}},\ Vol.~\bibinfo {volume} {53}\ (\bibinfo
  {publisher} {Elsevier},\ \bibinfo {year} {2009})\ pp.\ \bibinfo {pages}
  {293--363}\BibitemShut {NoStop}%
\bibitem [{\citenamefont {Gao}\ \emph {et~al.}(2014)\citenamefont {Gao},
  \citenamefont {Novitsky}, \citenamefont {Zhang}, \citenamefont {Cheong},
  \citenamefont {Gao}, \citenamefont {Lim}, \citenamefont {Luk'yanchuk},\ and\
  \citenamefont {Qiu}}]{Gao2014}%
  \BibitemOpen
  \bibfield  {author} {\bibinfo {author} {\bibfnamefont {D.}~\bibnamefont
  {Gao}}, \bibinfo {author} {\bibfnamefont {A.}~\bibnamefont {Novitsky}},
  \bibinfo {author} {\bibfnamefont {T.}~\bibnamefont {Zhang}}, \bibinfo
  {author} {\bibfnamefont {F.~C.}\ \bibnamefont {Cheong}}, \bibinfo {author}
  {\bibfnamefont {L.}~\bibnamefont {Gao}}, \bibinfo {author} {\bibfnamefont
  {C.~T.}\ \bibnamefont {Lim}}, \bibinfo {author} {\bibfnamefont
  {B.}~\bibnamefont {Luk'yanchuk}},\ and\ \bibinfo {author} {\bibfnamefont
  {C.-W.}\ \bibnamefont {Qiu}},\ }\href
  {https://doi.org/10.1002/lpor.201400071} {\bibfield  {journal} {\bibinfo
  {journal} {Laser {\&} Photonics Reviews}\ }\textbf {\bibinfo {volume} {9}},\
  \bibinfo {pages} {75} (\bibinfo {year} {2014})}\BibitemShut {NoStop}%
\bibitem [{\citenamefont {Yue}\ \emph {et~al.}(2019)\citenamefont {Yue},
  \citenamefont {Yan}, \citenamefont {Monks}, \citenamefont {Dhama},
  \citenamefont {Jiang}, \citenamefont {Minin}, \citenamefont {Minin},\ and\
  \citenamefont {Wang}}]{Yue2019}%
  \BibitemOpen
  \bibfield  {author} {\bibinfo {author} {\bibfnamefont {L.}~\bibnamefont
  {Yue}}, \bibinfo {author} {\bibfnamefont {B.}~\bibnamefont {Yan}}, \bibinfo
  {author} {\bibfnamefont {J.~N.}\ \bibnamefont {Monks}}, \bibinfo {author}
  {\bibfnamefont {R.}~\bibnamefont {Dhama}}, \bibinfo {author} {\bibfnamefont
  {C.}~\bibnamefont {Jiang}}, \bibinfo {author} {\bibfnamefont {O.~V.}\
  \bibnamefont {Minin}}, \bibinfo {author} {\bibfnamefont {I.~V.}\ \bibnamefont
  {Minin}},\ and\ \bibinfo {author} {\bibfnamefont {Z.}~\bibnamefont {Wang}},\
  }\href {https://doi.org/10.1038/s41598-019-56761-9} {\bibfield  {journal}
  {\bibinfo  {journal} {Scientific Reports}\ }\textbf {\bibinfo {volume} {9}},\
  \bibinfo {pages} {20224} (\bibinfo {year} {2019})}\BibitemShut {NoStop}%
\bibitem [{\citenamefont {Angelsky}\ \emph {et~al.}(2021)\citenamefont
  {Angelsky}, \citenamefont {Bekshaev}, \citenamefont {Hanson}, \citenamefont
  {Mokhun}, \citenamefont {Vasnetsov},\ and\ \citenamefont
  {Wang}}]{Angelsky2021}%
  \BibitemOpen
  \bibfield  {author} {\bibinfo {author} {\bibfnamefont {O.~V.}\ \bibnamefont
  {Angelsky}}, \bibinfo {author} {\bibfnamefont {A.~Y.}\ \bibnamefont
  {Bekshaev}}, \bibinfo {author} {\bibfnamefont {S.~G.}\ \bibnamefont
  {Hanson}}, \bibinfo {author} {\bibfnamefont {I.~I.}\ \bibnamefont {Mokhun}},
  \bibinfo {author} {\bibfnamefont {M.~V.}\ \bibnamefont {Vasnetsov}},\ and\
  \bibinfo {author} {\bibfnamefont {W.}~\bibnamefont {Wang}},\ }\href
  {https://doi.org/10.3389/fphy.2021.651964} {\bibfield  {journal} {\bibinfo
  {journal} {Frontiers in Physics}\ }\textbf {\bibinfo {volume} {9}},\ \bibinfo
  {pages} {651964} (\bibinfo {year} {2021})}\BibitemShut {NoStop}%
\bibitem [{\citenamefont {Gao}\ \emph {et~al.}(2017)\citenamefont {Gao},
  \citenamefont {Ding}, \citenamefont {Nieto-Vesperinas}, \citenamefont {Ding},
  \citenamefont {Rahman}, \citenamefont {Zhang}, \citenamefont {Lim},\ and\
  \citenamefont {Qiu}}]{Gao2017}%
  \BibitemOpen
  \bibfield  {author} {\bibinfo {author} {\bibfnamefont {D.}~\bibnamefont
  {Gao}}, \bibinfo {author} {\bibfnamefont {W.}~\bibnamefont {Ding}}, \bibinfo
  {author} {\bibfnamefont {M.}~\bibnamefont {Nieto-Vesperinas}}, \bibinfo
  {author} {\bibfnamefont {X.}~\bibnamefont {Ding}}, \bibinfo {author}
  {\bibfnamefont {M.}~\bibnamefont {Rahman}}, \bibinfo {author} {\bibfnamefont
  {T.}~\bibnamefont {Zhang}}, \bibinfo {author} {\bibfnamefont
  {C.}~\bibnamefont {Lim}},\ and\ \bibinfo {author} {\bibfnamefont {C.-W.}\
  \bibnamefont {Qiu}},\ }\href {https://doi.org/10.1038/lsa.2017.39} {\bibfield
   {journal} {\bibinfo  {journal} {Light: Science {\&} Applications}\ }\textbf
  {\bibinfo {volume} {6}},\ \bibinfo {pages} {e17039} (\bibinfo {year}
  {2017})}\BibitemShut {NoStop}%
\bibitem [{\citenamefont {Mokhun}\ \emph {et~al.}(2012)\citenamefont {Mokhun},
  \citenamefont {Arkhelyuk}, \citenamefont {Galushko}, \citenamefont
  {Kharitonovtta},\ and\ \citenamefont {Viktorovskaya}}]{Mokhun2012}%
  \BibitemOpen
  \bibfield  {author} {\bibinfo {author} {\bibfnamefont {I.}~\bibnamefont
  {Mokhun}}, \bibinfo {author} {\bibfnamefont {A.}~\bibnamefont {Arkhelyuk}},
  \bibinfo {author} {\bibfnamefont {Y.}~\bibnamefont {Galushko}}, \bibinfo
  {author} {\bibfnamefont {Y.}~\bibnamefont {Kharitonovtta}},\ and\ \bibinfo
  {author} {\bibfnamefont {J.}~\bibnamefont {Viktorovskaya}},\ }\href
  {https://doi.org/10.1364/AO.51.00C158} {\bibfield  {journal} {\bibinfo
  {journal} {Appl. Opt.}\ }\textbf {\bibinfo {volume} {51}},\ \bibinfo {pages}
  {C158} (\bibinfo {year} {2012})}\BibitemShut {NoStop}%
\bibitem [{\citenamefont {Bohren}(1983)}]{Bohren1983}%
  \BibitemOpen
  \bibfield  {author} {\bibinfo {author} {\bibfnamefont {C.~F.}\ \bibnamefont
  {Bohren}},\ }\href {https://doi.org/10.1119/1.13262} {\bibfield  {journal}
  {\bibinfo  {journal} {American Journal of Physics}\ }\textbf {\bibinfo
  {volume} {51}},\ \bibinfo {pages} {323} (\bibinfo {year} {1983})}\BibitemShut
  {NoStop}%
\bibitem [{\citenamefont {Wang}\ \emph {et~al.}(2004)\citenamefont {Wang},
  \citenamefont {Luk’yanchuk}, \citenamefont {Hong}, \citenamefont {Lin},\
  and\ \citenamefont {Chong}}]{Wang2004}%
  \BibitemOpen
  \bibfield  {author} {\bibinfo {author} {\bibfnamefont {Z.}~\bibnamefont
  {Wang}}, \bibinfo {author} {\bibfnamefont {B.}~\bibnamefont {Luk’yanchuk}},
  \bibinfo {author} {\bibfnamefont {M.}~\bibnamefont {Hong}}, \bibinfo {author}
  {\bibfnamefont {Y.}~\bibnamefont {Lin}},\ and\ \bibinfo {author}
  {\bibfnamefont {T.}~\bibnamefont {Chong}},\ }\href
  {https://doi.org/10.1103/PhysRevB.70.035418} {\bibfield  {journal} {\bibinfo
  {journal} {Physical Review B}\ }\textbf {\bibinfo {volume} {70}},\ \bibinfo
  {pages} {035418} (\bibinfo {year} {2004})}\BibitemShut {NoStop}%
\bibitem [{\citenamefont {Bashevoy}\ \emph {et~al.}(2005)\citenamefont
  {Bashevoy}, \citenamefont {Fedotov},\ and\ \citenamefont
  {Zheludev}}]{Bashevoy2005}%
  \BibitemOpen
  \bibfield  {author} {\bibinfo {author} {\bibfnamefont {M.}~\bibnamefont
  {Bashevoy}}, \bibinfo {author} {\bibfnamefont {V.}~\bibnamefont {Fedotov}},\
  and\ \bibinfo {author} {\bibfnamefont {N.}~\bibnamefont {Zheludev}},\ }\href
  {https://doi.org/10.1364/OPEX.13.008372} {\bibfield  {journal} {\bibinfo
  {journal} {Optics express}\ }\textbf {\bibinfo {volume} {13}},\ \bibinfo
  {pages} {8372} (\bibinfo {year} {2005})}\BibitemShut {NoStop}%
\bibitem [{\citenamefont {Tribelsky}\ and\ \citenamefont
  {Luk'yanchuk}(2006)}]{tribelsky2006anomalousPRL}%
  \BibitemOpen
  \bibfield  {author} {\bibinfo {author} {\bibfnamefont {M.~I.}\ \bibnamefont
  {Tribelsky}}\ and\ \bibinfo {author} {\bibfnamefont {B.~S.}\ \bibnamefont
  {Luk'yanchuk}},\ }\href
  {https://journals.aps.org/prl/abstract/10.1103/PhysRevLett.97.263902}
  {\bibfield  {journal} {\bibinfo  {journal} {Physical Review Letters}\
  }\textbf {\bibinfo {volume} {97}} (\bibinfo {year} {2006})}\BibitemShut
  {NoStop}%
\bibitem [{\citenamefont {Luk'yanchuk}\ \emph
  {et~al.}(2007{\natexlab{a}})\citenamefont {Luk'yanchuk}, \citenamefont
  {Wang}, \citenamefont {Tribelsky}, \citenamefont {Ternovsky}, \citenamefont
  {Hong},\ and\ \citenamefont {Chong}}]{luk2007peculiaritiesCOLA}%
  \BibitemOpen
  \bibfield  {author} {\bibinfo {author} {\bibfnamefont {B.}~\bibnamefont
  {Luk'yanchuk}}, \bibinfo {author} {\bibfnamefont {Z.}~\bibnamefont {Wang}},
  \bibinfo {author} {\bibfnamefont {M.}~\bibnamefont {Tribelsky}}, \bibinfo
  {author} {\bibfnamefont {V.}~\bibnamefont {Ternovsky}}, \bibinfo {author}
  {\bibfnamefont {M.}~\bibnamefont {Hong}},\ and\ \bibinfo {author}
  {\bibfnamefont {T.}~\bibnamefont {Chong}},\ }\href
  {https://iopscience.iop.org/article/10.1088/1742-6596/59/1/050/meta}
  {\bibfield  {journal} {\bibinfo  {journal} {Journal of Physics: Conference
  Series}\ }\textbf {\bibinfo {volume} {59}},\ \bibinfo {pages} {234} (\bibinfo
  {year} {2007}{\natexlab{a}})}\BibitemShut {NoStop}%
\bibitem [{\citenamefont {Luk'yanchuk}\ \emph
  {et~al.}(2007{\natexlab{b}})\citenamefont {Luk'yanchuk}, \citenamefont
  {Tribelsky}, \citenamefont {Ternovsky}, \citenamefont {Wang}, \citenamefont
  {Hong}, \citenamefont {Shi},\ and\ \citenamefont
  {Chong}}]{luk2007peculiaritiesJOA}%
  \BibitemOpen
  \bibfield  {author} {\bibinfo {author} {\bibfnamefont {B.}~\bibnamefont
  {Luk'yanchuk}}, \bibinfo {author} {\bibfnamefont {M.}~\bibnamefont
  {Tribelsky}}, \bibinfo {author} {\bibfnamefont {V.}~\bibnamefont
  {Ternovsky}}, \bibinfo {author} {\bibfnamefont {Z.}~\bibnamefont {Wang}},
  \bibinfo {author} {\bibfnamefont {M.}~\bibnamefont {Hong}}, \bibinfo {author}
  {\bibfnamefont {L.}~\bibnamefont {Shi}},\ and\ \bibinfo {author}
  {\bibfnamefont {T.}~\bibnamefont {Chong}},\ }\href
  {https://iopscience.iop.org/article/10.1088/1464-4258/9/9/S03/meta}
  {\bibfield  {journal} {\bibinfo  {journal} {Journal of Optics A: Pure and
  Applied Optics}\ }\textbf {\bibinfo {volume} {9}},\ \bibinfo {pages} {S294}
  (\bibinfo {year} {2007}{\natexlab{b}})}\BibitemShut {NoStop}%
\bibitem [{\citenamefont {Can\'{o}s}\ \emph {et~al.}(2021)\citenamefont
  {Can\'{o}s}, \citenamefont {Gurvitz}, \citenamefont {Benimetskiy},
  \citenamefont {Pidgayko}, \citenamefont {Samusev}, \citenamefont {Evlyukhin},
  \citenamefont {Bobrovs}, \citenamefont {Redka}, \citenamefont {Tribelsky},
  \citenamefont {Rahmani}, \citenamefont {Kamali}, \citenamefont {Pavlov},
  \citenamefont {Miroshnichenko},\ and\ \citenamefont
  {Shalin}}]{CanosValero2021}%
  \BibitemOpen
  \bibfield  {author} {\bibinfo {author} {\bibfnamefont {V.~A.}\ \bibnamefont
  {Can\'{o}s}}, \bibinfo {author} {\bibfnamefont {E.~A.}\ \bibnamefont
  {Gurvitz}}, \bibinfo {author} {\bibfnamefont {F.~A.}\ \bibnamefont
  {Benimetskiy}}, \bibinfo {author} {\bibfnamefont {D.~A.}\ \bibnamefont
  {Pidgayko}}, \bibinfo {author} {\bibfnamefont {A.}~\bibnamefont {Samusev}},
  \bibinfo {author} {\bibfnamefont {A.~B.}\ \bibnamefont {Evlyukhin}}, \bibinfo
  {author} {\bibfnamefont {V.}~\bibnamefont {Bobrovs}}, \bibinfo {author}
  {\bibfnamefont {D.}~\bibnamefont {Redka}}, \bibinfo {author} {\bibfnamefont
  {M.~I.}\ \bibnamefont {Tribelsky}}, \bibinfo {author} {\bibfnamefont
  {M.}~\bibnamefont {Rahmani}}, \bibinfo {author} {\bibfnamefont {K.~Z.}\
  \bibnamefont {Kamali}}, \bibinfo {author} {\bibfnamefont {A.~A.}\
  \bibnamefont {Pavlov}}, \bibinfo {author} {\bibfnamefont {A.~E.}\
  \bibnamefont {Miroshnichenko}},\ and\ \bibinfo {author} {\bibfnamefont
  {A.~S.}\ \bibnamefont {Shalin}},\ }\href
  {https://doi.org/https://doi.org/10.1002/lpor.202100114} {\bibfield
  {journal} {\bibinfo  {journal} {Laser \& Photonics Reviews}\ }\textbf
  {\bibinfo {volume} {15}},\ \bibinfo {pages} {2100114} (\bibinfo {year}
  {2021})}\BibitemShut {NoStop}%
\bibitem [{\citenamefont {Tribelsky}\ and\ \citenamefont
  {Miroshnichenko}(2022)}]{Tribel:2022_UFN}%
  \BibitemOpen
  \bibfield  {author} {\bibinfo {author} {\bibfnamefont {M.~I.}\ \bibnamefont
  {Tribelsky}}\ and\ \bibinfo {author} {\bibfnamefont {A.~E.}\ \bibnamefont
  {Miroshnichenko}},\ }\href@noop {} {\bibfield  {journal} {\bibinfo  {journal}
  {Physics-Uspekhi}\ }\textbf {\bibinfo {volume} {65}},\ \bibinfo {pages} {40}
  (\bibinfo {year} {2022})}\BibitemShut {NoStop}%
\bibitem [{\citenamefont {Yue}\ \emph {et~al.}(2022)\citenamefont {Yue},
  \citenamefont {Wang}, \citenamefont {Yan}, \citenamefont {Xie}, \citenamefont
  {Geints}, \citenamefont {Minin},\ and\ \citenamefont {Minin}}]{Yue2022}%
  \BibitemOpen
  \bibfield  {author} {\bibinfo {author} {\bibfnamefont {L.}~\bibnamefont
  {Yue}}, \bibinfo {author} {\bibfnamefont {Z.}~\bibnamefont {Wang}}, \bibinfo
  {author} {\bibfnamefont {B.}~\bibnamefont {Yan}}, \bibinfo {author}
  {\bibfnamefont {Y.}~\bibnamefont {Xie}}, \bibinfo {author} {\bibfnamefont
  {Y.~E.}\ \bibnamefont {Geints}}, \bibinfo {author} {\bibfnamefont {O.~V.}\
  \bibnamefont {Minin}},\ and\ \bibinfo {author} {\bibfnamefont {I.~V.}\
  \bibnamefont {Minin}},\ }\href {https://doi.org/10.3390/photonics9030154}
  {\bibfield  {journal} {\bibinfo  {journal} {Photonics}\ }\textbf {\bibinfo
  {volume} {9}},\ \bibinfo {pages} {154} (\bibinfo {year} {2022})}\BibitemShut
  {NoStop}%
\bibitem [{\citenamefont {Geints}\ \emph
  {et~al.}(2022{\natexlab{a}})\citenamefont {Geints}, \citenamefont {Minin},\
  and\ \citenamefont {Minin}}]{Geints2022}%
  \BibitemOpen
  \bibfield  {author} {\bibinfo {author} {\bibfnamefont {Y.~E.}\ \bibnamefont
  {Geints}}, \bibinfo {author} {\bibfnamefont {I.~V.}\ \bibnamefont {Minin}},\
  and\ \bibinfo {author} {\bibfnamefont {O.~V.}\ \bibnamefont {Minin}},\ }\href
  {https://doi.org/10.1364/ol.452683} {\bibfield  {journal} {\bibinfo
  {journal} {Optics Letters}\ }\textbf {\bibinfo {volume} {47}},\ \bibinfo
  {pages} {1786} (\bibinfo {year} {2022}{\natexlab{a}})}\BibitemShut {NoStop}%
\bibitem [{\citenamefont {Geints}\ \emph
  {et~al.}(2022{\natexlab{b}})\citenamefont {Geints}, \citenamefont {Minin},\
  and\ \citenamefont {Minin}}]{Geints2022a}%
  \BibitemOpen
  \bibfield  {author} {\bibinfo {author} {\bibfnamefont {Y.~E.}\ \bibnamefont
  {Geints}}, \bibinfo {author} {\bibfnamefont {I.~V.}\ \bibnamefont {Minin}},\
  and\ \bibinfo {author} {\bibfnamefont {O.~V.}\ \bibnamefont {Minin}},\ }\href
  {https://doi.org/10.1016/j.optcom.2022.128779} {\bibfield  {journal}
  {\bibinfo  {journal} {Optics Communications}\ }\textbf {\bibinfo {volume}
  {524}},\ \bibinfo {pages} {128779} (\bibinfo {year}
  {2022}{\natexlab{b}})}\BibitemShut {NoStop}%
\bibitem [{\citenamefont {Berry}\ \emph {et~al.}(2019)\citenamefont {Berry},
  \citenamefont {Zheludev}, \citenamefont {Aharonov}, \citenamefont {Colombo},
  \citenamefont {Sabadini}, \citenamefont {Struppa}, \citenamefont {Tollaksen},
  \citenamefont {Rogers}, \citenamefont {Qin}, \citenamefont {Hong},
  \citenamefont {Luo}, \citenamefont {Remez}, \citenamefont {Arie},
  \citenamefont {Götte}, \citenamefont {Dennis}, \citenamefont {Wong},
  \citenamefont {Eleftheriades}, \citenamefont {Eliezer}, \citenamefont
  {Bahabad}, \citenamefont {Chen}, \citenamefont {Wen}, \citenamefont {Liang},
  \citenamefont {Hao}, \citenamefont {Qiu}, \citenamefont {Kempf},
  \citenamefont {Katzav},\ and\ \citenamefont {Schwartz}}]{Berry2019}%
  \BibitemOpen
  \bibfield  {author} {\bibinfo {author} {\bibfnamefont {M.}~\bibnamefont
  {Berry}}, \bibinfo {author} {\bibfnamefont {N.}~\bibnamefont {Zheludev}},
  \bibinfo {author} {\bibfnamefont {Y.}~\bibnamefont {Aharonov}}, \bibinfo
  {author} {\bibfnamefont {F.}~\bibnamefont {Colombo}}, \bibinfo {author}
  {\bibfnamefont {I.}~\bibnamefont {Sabadini}}, \bibinfo {author}
  {\bibfnamefont {D.~C.}\ \bibnamefont {Struppa}}, \bibinfo {author}
  {\bibfnamefont {J.}~\bibnamefont {Tollaksen}}, \bibinfo {author}
  {\bibfnamefont {E.~T.~F.}\ \bibnamefont {Rogers}}, \bibinfo {author}
  {\bibfnamefont {F.}~\bibnamefont {Qin}}, \bibinfo {author} {\bibfnamefont
  {M.}~\bibnamefont {Hong}}, \bibinfo {author} {\bibfnamefont {X.}~\bibnamefont
  {Luo}}, \bibinfo {author} {\bibfnamefont {R.}~\bibnamefont {Remez}}, \bibinfo
  {author} {\bibfnamefont {A.}~\bibnamefont {Arie}}, \bibinfo {author}
  {\bibfnamefont {J.~B.}\ \bibnamefont {Götte}}, \bibinfo {author}
  {\bibfnamefont {M.~R.}\ \bibnamefont {Dennis}}, \bibinfo {author}
  {\bibfnamefont {A.~M.~H.}\ \bibnamefont {Wong}}, \bibinfo {author}
  {\bibfnamefont {G.~V.}\ \bibnamefont {Eleftheriades}}, \bibinfo {author}
  {\bibfnamefont {Y.}~\bibnamefont {Eliezer}}, \bibinfo {author} {\bibfnamefont
  {A.}~\bibnamefont {Bahabad}}, \bibinfo {author} {\bibfnamefont
  {G.}~\bibnamefont {Chen}}, \bibinfo {author} {\bibfnamefont {Z.}~\bibnamefont
  {Wen}}, \bibinfo {author} {\bibfnamefont {G.}~\bibnamefont {Liang}}, \bibinfo
  {author} {\bibfnamefont {C.}~\bibnamefont {Hao}}, \bibinfo {author}
  {\bibfnamefont {C.-W.}\ \bibnamefont {Qiu}}, \bibinfo {author} {\bibfnamefont
  {A.}~\bibnamefont {Kempf}}, \bibinfo {author} {\bibfnamefont
  {E.}~\bibnamefont {Katzav}},\ and\ \bibinfo {author} {\bibfnamefont
  {M.}~\bibnamefont {Schwartz}},\ }\href
  {https://doi.org/10.1088/2040-8986/ab0191} {\bibfield  {journal} {\bibinfo
  {journal} {Journal of Optics}\ }\textbf {\bibinfo {volume} {21}},\ \bibinfo
  {pages} {053002} (\bibinfo {year} {2019})}\BibitemShut {NoStop}%
\bibitem [{\citenamefont {Ospanova}\ \emph {et~al.}(2021)\citenamefont
  {Ospanova}, \citenamefont {Basharin}, \citenamefont {Miroshnichenko},\ and\
  \citenamefont {Luk’yanchuk}}]{ospanova2021generalized}%
  \BibitemOpen
  \bibfield  {author} {\bibinfo {author} {\bibfnamefont {A.~K.}\ \bibnamefont
  {Ospanova}}, \bibinfo {author} {\bibfnamefont {A.}~\bibnamefont {Basharin}},
  \bibinfo {author} {\bibfnamefont {A.~E.}\ \bibnamefont {Miroshnichenko}},\
  and\ \bibinfo {author} {\bibfnamefont {B.}~\bibnamefont {Luk’yanchuk}},\
  }\href {https://opg.optica.org/ome/fulltext.cfm?uri=ome-11-1-23&id=444227}
  {\bibfield  {journal} {\bibinfo  {journal} {Optical Materials Express}\
  }\textbf {\bibinfo {volume} {11}},\ \bibinfo {pages} {23} (\bibinfo {year}
  {2021})}\BibitemShut {NoStop}%
\bibitem [{\citenamefont {Kotlyar}\ \emph {et~al.}(2018)\citenamefont
  {Kotlyar}, \citenamefont {Nalimov},\ and\ \citenamefont
  {Stafeev}}]{kotliar2018obratnyi}%
  \BibitemOpen
  \bibfield  {author} {\bibinfo {author} {\bibfnamefont {V.~V.}\ \bibnamefont
  {Kotlyar}}, \bibinfo {author} {\bibfnamefont {A.~G.}\ \bibnamefont
  {Nalimov}},\ and\ \bibinfo {author} {\bibfnamefont {S.~S.}\ \bibnamefont
  {Stafeev}},\ }\href
  {https://cyberleninka.ru/article/n/obratnyy-potok-energii-vblizi-opticheskoy-osi-v-oblasti-ostrogo-fokusa-opticheskogo-vihrya-s-krugovoy-polyarizatsiey}
  {\bibfield  {journal} {\bibinfo  {journal} {Komp'yuternaya optika (Computer
  Optics; in Russian)}\ }\textbf {\bibinfo {volume} {42}},\ \bibinfo {pages}
  {392} (\bibinfo {year} {2018})}\BibitemShut {NoStop}%
\bibitem [{\citenamefont {Kotlyar}\ \emph {et~al.}(2019)\citenamefont
  {Kotlyar}, \citenamefont {Stafeev},\ and\ \citenamefont
  {Kovalev}}]{Kotlyar2019}%
  \BibitemOpen
  \bibfield  {author} {\bibinfo {author} {\bibfnamefont {V.~V.}\ \bibnamefont
  {Kotlyar}}, \bibinfo {author} {\bibfnamefont {S.~S.}\ \bibnamefont
  {Stafeev}},\ and\ \bibinfo {author} {\bibfnamefont {A.~A.}\ \bibnamefont
  {Kovalev}},\ }\href {https://doi.org/10.1364/oe.27.016689} {\bibfield
  {journal} {\bibinfo  {journal} {Optics Express}\ }\textbf {\bibinfo {volume}
  {27}},\ \bibinfo {pages} {16689} (\bibinfo {year} {2019})}\BibitemShut
  {NoStop}%
\bibitem [{\citenamefont {Berry}(2004)}]{Berry2004_TC}%
  \BibitemOpen
  \bibfield  {author} {\bibinfo {author} {\bibfnamefont {M.~V.}\ \bibnamefont
  {Berry}},\ }\href {https://doi.org/10.1088/1464-4258/6/2/018} {\bibfield
  {journal} {\bibinfo  {journal} {Journal of Optics A: Pure and Applied
  Optics}\ }\textbf {\bibinfo {volume} {6}},\ \bibinfo {pages} {259} (\bibinfo
  {year} {2004})}\BibitemShut {NoStop}%
\bibitem [{\citenamefont {Kotlyar}\ and\ \citenamefont
  {Kovalev}(2021)}]{kotlyiarBookTC}%
  \BibitemOpen
  \bibfield  {author} {\bibinfo {author} {\bibfnamefont {V.~V.}\ \bibnamefont
  {Kotlyar}}\ and\ \bibinfo {author} {\bibfnamefont {A.~A.}\ \bibnamefont
  {Kovalev}},\ }\href@noop {} {\emph {\bibinfo {title} {Topologicheskii zaryad
  opticheskikh vikhrei (Topological Charge of Optical Votices; in Russian)}}}\
  (\bibinfo  {publisher} {OOO ``Novaya tekhnika'', Samara},\ \bibinfo {year}
  {2021})\BibitemShut {NoStop}%
\bibitem [{\citenamefont {Berry}(1984)}]{berry1984quantal}%
  \BibitemOpen
  \bibfield  {author} {\bibinfo {author} {\bibfnamefont {M.~V.}\ \bibnamefont
  {Berry}},\ }\href
  {https://royalsocietypublishing.org/doi/abs/10.1098/rspa.1984.0023}
  {\bibfield  {journal} {\bibinfo  {journal} {Proceedings of the Royal Society
  of London. A. Mathematical and Physical Sciences}\ }\textbf {\bibinfo
  {volume} {392}},\ \bibinfo {pages} {45} (\bibinfo {year} {1984})}\BibitemShut
  {NoStop}%
\bibitem [{\citenamefont {Cohen}\ \emph {et~al.}(2019)\citenamefont {Cohen},
  \citenamefont {Larocque}, \citenamefont {Bouchard}, \citenamefont
  {Nejadsattari}, \citenamefont {Gefen},\ and\ \citenamefont
  {Karimi}}]{Cohen2019}%
  \BibitemOpen
  \bibfield  {author} {\bibinfo {author} {\bibfnamefont {E.}~\bibnamefont
  {Cohen}}, \bibinfo {author} {\bibfnamefont {H.}~\bibnamefont {Larocque}},
  \bibinfo {author} {\bibfnamefont {F.}~\bibnamefont {Bouchard}}, \bibinfo
  {author} {\bibfnamefont {F.}~\bibnamefont {Nejadsattari}}, \bibinfo {author}
  {\bibfnamefont {Y.}~\bibnamefont {Gefen}},\ and\ \bibinfo {author}
  {\bibfnamefont {E.}~\bibnamefont {Karimi}},\ }\href
  {https://doi.org/10.1038/s42254-019-0071-1} {\bibfield  {journal} {\bibinfo
  {journal} {Nature Reviews Physics}\ }\textbf {\bibinfo {volume} {1}},\
  \bibinfo {pages} {437} (\bibinfo {year} {2019})}\BibitemShut {NoStop}%
\bibitem [{\citenamefont {Tribelsky}\ and\ \citenamefont
  {Rubinstein}(2022{\natexlab{a}})}]{Tribelsky2022}%
  \BibitemOpen
  \bibfield  {author} {\bibinfo {author} {\bibfnamefont {M.~I.}\ \bibnamefont
  {Tribelsky}}\ and\ \bibinfo {author} {\bibfnamefont {B.~Y.}\ \bibnamefont
  {Rubinstein}},\ }\href {https://doi.org/10.3390/nano12111878} {\bibfield
  {journal} {\bibinfo  {journal} {Nanomaterials}\ }\textbf {\bibinfo {volume}
  {12}},\ \bibinfo {pages} {1878} (\bibinfo {year}
  {2022}{\natexlab{a}})}\BibitemShut {NoStop}%
\bibitem [{\citenamefont {Tribelsky}\ and\ \citenamefont
  {Rubinstein}(2022{\natexlab{b}})}]{Tribelsky2022_Nanomat_diss}%
  \BibitemOpen
  \bibfield  {author} {\bibinfo {author} {\bibfnamefont {M.~I.}\ \bibnamefont
  {Tribelsky}}\ and\ \bibinfo {author} {\bibfnamefont {B.~Y.}\ \bibnamefont
  {Rubinstein}},\ }\href {https://doi.org/10.3390/nano12183164} {\bibfield
  {journal} {\bibinfo  {journal} {Nanomaterials}\ }\textbf {\bibinfo {volume}
  {12}},\ \bibinfo {pages} {3164} (\bibinfo {year}
  {2022}{\natexlab{b}})}\BibitemShut {NoStop}%
\bibitem [{\citenamefont {Tribelsky}(2023)}]{Tribel:2023_arXiv}%
  \BibitemOpen
  \bibfield  {author} {\bibinfo {author} {\bibfnamefont {M.~I.}\ \bibnamefont
  {Tribelsky}},\ }\href@noop {} {\bibfield  {journal} {\bibinfo  {journal}
  {arXiv.org}\ } (\bibinfo {year} {2023})},\ \Eprint
  {https://arxiv.org/abs/2305.08534v2} {2305.08534v2} \BibitemShut {NoStop}%
\bibitem [{\citenamefont {Bohren}\ and\ \citenamefont
  {Huffman}(1998)}]{Bohren1998}%
  \BibitemOpen
  \bibfield  {author} {\bibinfo {author} {\bibfnamefont {C.~F.}\ \bibnamefont
  {Bohren}}\ and\ \bibinfo {author} {\bibfnamefont {D.~R.}\ \bibnamefont
  {Huffman}},\ }\href@noop {} {\emph {\bibinfo {title} {Absorption and
  Scattering of Light by Small Particles}}}\ (\bibinfo  {publisher} {WILEY-VCH
  Verlag, New York},\ \bibinfo {year} {1998})\BibitemShut {NoStop}%
\bibitem [{\citenamefont {Landau}\ \emph {et~al.}(2013)\citenamefont {Landau},
  \citenamefont {Bell}, \citenamefont {Kearsley}, \citenamefont {Pitaevskii},
  \citenamefont {Lifshitz},\ and\ \citenamefont {Sykes}}]{Landau_Electrodyn}%
  \BibitemOpen
  \bibfield  {author} {\bibinfo {author} {\bibfnamefont {L.~D.}\ \bibnamefont
  {Landau}}, \bibinfo {author} {\bibfnamefont {J.~S.}\ \bibnamefont {Bell}},
  \bibinfo {author} {\bibfnamefont {M.}~\bibnamefont {Kearsley}}, \bibinfo
  {author} {\bibfnamefont {L.}~\bibnamefont {Pitaevskii}}, \bibinfo {author}
  {\bibfnamefont {E.}~\bibnamefont {Lifshitz}},\ and\ \bibinfo {author}
  {\bibfnamefont {J.}~\bibnamefont {Sykes}},\ }\href@noop {} {\emph {\bibinfo
  {title} {Electrodynamics of continuous media}}},\ Vol.~\bibinfo {volume} {8}\
  (\bibinfo  {publisher} {Elsevier, Amsterdam},\ \bibinfo {year}
  {2013})\BibitemShut {NoStop}%
\bibitem [{\citenamefont {Jackson}(1998)}]{Jackson1998}%
  \BibitemOpen
  \bibfield  {author} {\bibinfo {author} {\bibfnamefont {J.~D.}\ \bibnamefont
  {Jackson}},\ }\href@noop {} {\emph {\bibinfo {title} {Classical
  Electrodynamics}}}\ (\bibinfo  {publisher} {Wiley, New York},\ \bibinfo
  {year} {1998})\ p.\ \bibinfo {pages} {832}\BibitemShut {NoStop}%
\bibitem [{\citenamefont {Bliokh}\ \emph
  {et~al.}(2014{\natexlab{a}})\citenamefont {Bliokh}, \citenamefont {Kivshar},\
  and\ \citenamefont {Nori}}]{bliokh2014magnetoelectric}%
  \BibitemOpen
  \bibfield  {author} {\bibinfo {author} {\bibfnamefont {K.~Y.}\ \bibnamefont
  {Bliokh}}, \bibinfo {author} {\bibfnamefont {Y.~S.}\ \bibnamefont
  {Kivshar}},\ and\ \bibinfo {author} {\bibfnamefont {F.}~\bibnamefont
  {Nori}},\ }\href
  {https://journals.aps.org/prl/abstract/10.1103/PhysRevLett.113.033601}
  {\bibfield  {journal} {\bibinfo  {journal} {Physical review letters}\
  }\textbf {\bibinfo {volume} {113}},\ \bibinfo {pages} {033601} (\bibinfo
  {year} {2014}{\natexlab{a}})}\BibitemShut {NoStop}%
\bibitem [{\citenamefont {Bliokh}\ \emph
  {et~al.}(2014{\natexlab{b}})\citenamefont {Bliokh}, \citenamefont
  {Bekshaev},\ and\ \citenamefont {Nori}}]{Bliokh2014}%
  \BibitemOpen
  \bibfield  {author} {\bibinfo {author} {\bibfnamefont {K.~Y.}\ \bibnamefont
  {Bliokh}}, \bibinfo {author} {\bibfnamefont {A.~Y.}\ \bibnamefont
  {Bekshaev}},\ and\ \bibinfo {author} {\bibfnamefont {F.}~\bibnamefont
  {Nori}},\ }\bibfield  {journal} {\bibinfo  {journal} {Nature Communications}\
  }\textbf {\bibinfo {volume} {5}},\ \href {https://doi.org/10.1038/ncomms4300}
  {10.1038/ncomms4300} (\bibinfo {year} {2014}{\natexlab{b}})\BibitemShut
  {NoStop}%
\bibitem [{\citenamefont {Bekshaev}\ \emph {et~al.}(2015)\citenamefont
  {Bekshaev}, \citenamefont {Bliokh},\ and\ \citenamefont
  {Nori}}]{Bekshaev2015}%
  \BibitemOpen
  \bibfield  {author} {\bibinfo {author} {\bibfnamefont {A.~Y.}\ \bibnamefont
  {Bekshaev}}, \bibinfo {author} {\bibfnamefont {K.~Y.}\ \bibnamefont
  {Bliokh}},\ and\ \bibinfo {author} {\bibfnamefont {F.}~\bibnamefont {Nori}},\
  }\href {https://doi.org/10.1103/physrevx.5.011039} {\bibfield  {journal}
  {\bibinfo  {journal} {Physical Review X}\ }\textbf {\bibinfo {volume} {5}},\
  \bibinfo {pages} {011039} (\bibinfo {year} {2015})}\BibitemShut {NoStop}%
\bibitem [{\citenamefont {Xu}\ and\ \citenamefont
  {Nieto-Vesperinas}(2019)}]{Xu2019}%
  \BibitemOpen
  \bibfield  {author} {\bibinfo {author} {\bibfnamefont {X.}~\bibnamefont
  {Xu}}\ and\ \bibinfo {author} {\bibfnamefont {M.}~\bibnamefont
  {Nieto-Vesperinas}},\ }\href
  {https://journals.aps.org/prl/abstract/10.1103/PhysRevLett.123.233902}
  {\bibfield  {journal} {\bibinfo  {journal} {Physical review letters}\
  }\textbf {\bibinfo {volume} {123}},\ \bibinfo {pages} {233902} (\bibinfo
  {year} {2019})}\BibitemShut {NoStop}%
\bibitem [{\citenamefont {Khonina}\ \emph {et~al.}(2021)\citenamefont
  {Khonina}, \citenamefont {Degtyarev}, \citenamefont {Ustinov},\ and\
  \citenamefont {Porfirev}}]{Khonina2021}%
  \BibitemOpen
  \bibfield  {author} {\bibinfo {author} {\bibfnamefont {S.~N.}\ \bibnamefont
  {Khonina}}, \bibinfo {author} {\bibfnamefont {S.~A.}\ \bibnamefont
  {Degtyarev}}, \bibinfo {author} {\bibfnamefont {A.~V.}\ \bibnamefont
  {Ustinov}},\ and\ \bibinfo {author} {\bibfnamefont {A.~P.}\ \bibnamefont
  {Porfirev}},\ }\href {https://doi.org/10.1364/oe.428453} {\bibfield
  {journal} {\bibinfo  {journal} {Optics Express}\ }\textbf {\bibinfo {volume}
  {29}},\ \bibinfo {pages} {18634} (\bibinfo {year} {2021})}\BibitemShut
  {NoStop}%
\bibitem [{\citenamefont {Tang}\ and\ \citenamefont {Cohen}(2010)}]{Tang2010}%
  \BibitemOpen
  \bibfield  {author} {\bibinfo {author} {\bibfnamefont {Y.}~\bibnamefont
  {Tang}}\ and\ \bibinfo {author} {\bibfnamefont {A.~E.}\ \bibnamefont
  {Cohen}},\ }\href {https://doi.org/10.1103/physrevlett.104.163901} {\bibfield
   {journal} {\bibinfo  {journal} {Physical Review Letters}\ }\textbf {\bibinfo
  {volume} {104}},\ \bibinfo {pages} {163901} (\bibinfo {year}
  {2010})}\BibitemShut {NoStop}%
\bibitem [{\citenamefont {Lininger}\ \emph {et~al.}(2022)\citenamefont
  {Lininger}, \citenamefont {Palermo}, \citenamefont {Guglielmelli},
  \citenamefont {Nicoletta}, \citenamefont {Goel}, \citenamefont {Hinczewski},\
  and\ \citenamefont {Strangi}}]{Lininger2022}%
  \BibitemOpen
  \bibfield  {author} {\bibinfo {author} {\bibfnamefont {A.}~\bibnamefont
  {Lininger}}, \bibinfo {author} {\bibfnamefont {G.}~\bibnamefont {Palermo}},
  \bibinfo {author} {\bibfnamefont {A.}~\bibnamefont {Guglielmelli}}, \bibinfo
  {author} {\bibfnamefont {G.}~\bibnamefont {Nicoletta}}, \bibinfo {author}
  {\bibfnamefont {M.}~\bibnamefont {Goel}}, \bibinfo {author} {\bibfnamefont
  {M.}~\bibnamefont {Hinczewski}},\ and\ \bibinfo {author} {\bibfnamefont
  {G.}~\bibnamefont {Strangi}},\ }\href
  {https://doi.org/10.1002/adma.202107325} {\bibfield  {journal} {\bibinfo
  {journal} {Advanced Materials}\ ,\ \bibinfo {pages} {2107325}} (\bibinfo
  {year} {2022})}\BibitemShut {NoStop}%
\bibitem [{\citenamefont {Arnold}(2012)}]{ODE:Arnold}%
  \BibitemOpen
  \bibfield  {author} {\bibinfo {author} {\bibfnamefont {V.~I.}\ \bibnamefont
  {Arnold}},\ }\href@noop {} {\emph {\bibinfo {title} {Obyknovennye
  differentcialnye uravneniya (Ordinary Differential Equations; in Russian)}}}\
  (\bibinfo  {publisher} {MTCNMO, Moscow},\ \bibinfo {year} {2012})\BibitemShut
  {NoStop}%
\bibitem [{\citenamefont {Polyanskiy}()}]{Polyanskiy}%
  \BibitemOpen
  \bibfield  {author} {\bibinfo {author} {\bibfnamefont {M.}~\bibnamefont
  {Polyanskiy}},\ }\href {http://refractiveindex.info/} {\bibinfo {title}
  {Refractive index database}},\ \bibinfo {howpublished}
  {\url{http://refractiveindex.info/}},\ \bibinfo {note} {accessed: May. 25,
  2022}\BibitemShut {NoStop}%
\bibitem [{\citenamefont {Bekshaev}\ and\ \citenamefont
  {Soskin}(2007)}]{bekshaev2007transverse}%
  \BibitemOpen
  \bibfield  {author} {\bibinfo {author} {\bibfnamefont {A.~Y.}\ \bibnamefont
  {Bekshaev}}\ and\ \bibinfo {author} {\bibfnamefont {M.}~\bibnamefont
  {Soskin}},\ }\href
  {https://www.sciencedirect.com/science/article/abs/pii/S0030401806011618}
  {\bibfield  {journal} {\bibinfo  {journal} {Optics communications}\ }\textbf
  {\bibinfo {volume} {271}},\ \bibinfo {pages} {332} (\bibinfo {year}
  {2007})}\BibitemShut {NoStop}%
\bibitem [{\citenamefont {Bekshaev}\ \emph {et~al.}(2011)\citenamefont
  {Bekshaev}, \citenamefont {Bliokh},\ and\ \citenamefont
  {Soskin}}]{bekshaev2011internal}%
  \BibitemOpen
  \bibfield  {author} {\bibinfo {author} {\bibfnamefont {A.}~\bibnamefont
  {Bekshaev}}, \bibinfo {author} {\bibfnamefont {K.~Y.}\ \bibnamefont
  {Bliokh}},\ and\ \bibinfo {author} {\bibfnamefont {M.}~\bibnamefont
  {Soskin}},\ }\href
  {https://iopscience.iop.org/article/10.1088/2040-8978/13/5/053001} {\bibfield
   {journal} {\bibinfo  {journal} {Journal of Optics}\ }\textbf {\bibinfo
  {volume} {13}},\ \bibinfo {pages} {053001} (\bibinfo {year}
  {2011})}\BibitemShut {NoStop}%
\end{thebibliography}%
\end{document}